\documentclass[useAMS,usenatbib]{mn2e}

\usepackage{graphicx,amssymb,color}
\usepackage[normalem]{ulem}

\newcommand{\rev}{  }

\title[Impact crater ejecta in WD systems]
{Generating metal-polluting debris in white dwarf planetary systems from small-impact crater ejecta}

\author[Veras \& Kurosawa]{
Dimitri Veras$^{1,2}$\thanks{E-mail: d.veras@warwick.ac.uk}\thanks{STFC Ernest Rutherford Fellow},
Kosuke Kurosawa$^{3}$
\\
$^{1}$Centre for Exoplanets and Habitability, University of Warwick, Coventry CV4 7AL, UK
\\
$^{2}$Department of Physics, University of Warwick, Coventry CV4 7AL, UK
\\
$^{3}$Planetary Exploration Research Center, Chiba Institute of Technology, 2-17-1, Narashino, Tsudanuma, 
Chiba 275-0016, Japan
}

\pubyear{2019}

\begin{document}
\label{firstpage}
\pagerange{\pageref{firstpage}--\pageref{lastpage}}
\maketitle

\begin{abstract}
Metal pollution in white dwarf photospheres originates from the accretion of some combination of planets, moons, asteroids, comets, boulders, pebbles and dust. When large bodies reside in dynamically stagnant locations -- unable themselves to pollute nor even closely approach the white dwarf -- then smaller reservoirs of impact debris may become a complementary or the primary source of metal pollutants. Here, we take a first step towards exploring this possibility by computing limits on the recoil mass that escapes the gravitational pull of the target object following a single impact onto an atmosphere-less surface. By considering vertical impacts only with the full-chain analytical prescription from \cite{kurtak2019}, we provide lower bounds for the ejected mass for basalt, granite, iron and water-rich target objects across the radii range $10^{0-3}$ km. Our use of the full-chain prescription as opposed to physical experiments or hydrocode simulations allows us to quickly sample a wide range of parameter space appropriate to white dwarf planetary systems. Our numerical results could be used in future studies to constrain freshly-generated small debris reservoirs around white dwarfs given a particular planetary system architecture, bombardment history, and impact geometries.
\end{abstract}


\section{Introduction}

\subsection{Fates of planetary systems}

As the endpoint of stellar evolution for the majority of stars in the Galaxy, white dwarfs showcase the fate of planetary systems. Our Sun will become a white dwarf, and will likely host five or six major planets that will have managed to survive the intermediate giant branch phase of Solar evolution \citep{schcon2008,veras2016a,schetal2019}. The Solar white dwarf will also host moons, comets, asteroids, boulders, pebbles and dust.

Radiative forces and dynamical instability amongst these objects could lead some to migrate towards and accrete onto the solar white dwarf \citep{veras2016b}. Because white dwarfs have surface gravities which are $10^5$ higher than the Earth's, white dwarf atmospheres chemically stratify this accreted matter \citep{schatzman1958,paqetal1986,koester2009}.

\subsubsection{Chemistry of white dwarf metal pollution}

Because white dwarf atmospheres are composed exclusively of hydrogen, helium or some combination of the two, accreted planetary material is easily distinguished\footnote{Any fallback material from the progenitor's stellar winds would sink at too early of an epoch to be observed and is anyway compositionally dis-similar from accreted planetary material.}, and regularly observed from both the ground (\citealt*{dufetal2007}, \citealt*{kleetal2010,kleetal2011}, \citealt*{kleetal2013}, \citealt*{wiletal2014}, \citealt*{genetal2015}, \citealt*{kepetal2015,kepetal2016}, \citealt*{holetal2017,holetal2018}) and from space (\citealt*{gaeetal2012}, \citealt*{juretal2012}, \citealt*{xuetal2013,xuetal2014}, \citealt*{wiletal2015,wiletal2016}, \citealt*{melduf2017}, \citealt*{couetal2019}). As a result, we can extract the bulk chemical composition of exo-planetary bodies by observing their broken-up remains \citep{juryou2014}. These autopsies represent the only direct currently available means to probe rocky interiors.

The above studies and others have so far yielded the detection of the following 20 metals (with atomic number): C(6), N(7), O(8), Na(11), Mg(12), Al(13), Si(14), P(15), S(16), Ca(20), Sc(21), Ti(22), V(23), Cr(24), Mn(25), Fe(26), Co(27), Ni(28), Cu(29) and Sr(38). Nearly all polluted white dwarfs contain either Ca or Mg, which generate the strongest spectral signatures; in contrast, for only one white dwarf has N been detected \citep{xuetal2017}. Some white dwarfs contain at least ten different observable exoplanetary metals \citep[e.g.][]{dufetal2012,melduf2017,swaetal2019,xuetal2019}, and about two dozen white dwarfs currently host at least five metals \citep{haretal2018,holetal2018}. In each case, because of spectroscopic limitations, we obtain only a partial snapshot of the bulk chemical composition of progenitor exoplanetary material.

Overall, the ensemble of metals found in white dwarf atmospheres is indicative of dry (volatile-poor) progenitors with compositions similar to the bulk Earth {\rev \citep{gaeetal2012,juryou2014,haretal2018,holetal2018,zucyou2018,doyetal2019,bonetal2020}. Determining if the composition of the surrounding debris and gas matches that of the metal pollution is challenging\footnote{{\rev Around younger, main-sequence stars, IR observations have revolutionized the fields of planet formation (see reviews by \citealt*{wyatt2008}, \citealt*{kraetal2018} and \citealt*{andrews2020}), although determining grain composition from these observations remains nuanced.} }, and so far has yielded constraints in just a few cases: for the close and bright G29-38 \citep{reaetal2005,reaetal2009}, the minor planet host WD 1145+017 \citep{xuetal2016,cauetal2018,foretal2020} and the major planet host WD J0914+1914 \citep{ganetal2019}. Nevertheless, the above evidence for volatile-poor atmospheric pollution is overwhelming.} Therefore in this paper, the consideration of a terrestrial-like composition (at least with granite or basalt) is paramount. 

Some white dwarfs, however, showcase high-profile exceptions. In a few instances, the progenitors were volatile-rich and were composed of water by mass fractions of tens of per cent \citep{faretal2013,radetal2015,genetal2017,xuetal2017}. These observations are supported by theoretical studies which have quantified how internal water can be retained even during the giant branch phases of stellar evolution \citep{malper2016,malper2017a,malper2017b}.  Hence, in this paper we also consider water-rich bodies. Finally, one polluted white dwarf was recently shown to be orbited by a ferrous core fragment \citep{manetal2019}, indicating the importance of modelling iron-rich bodies, which we also consider here.

We therefore choose basalt, granite, water and iron in this paper as representations of exoplanetary material solely based on white dwarf metal pollution. White dwarf pollution provides the most direct way of measuring exoplanetary bulk chemical compositions, because observational data on major exoplanets is, at best, otherwise largely restricted to their masses and radii only. Although a handful of chemical elements and molecules have been positively detected in main-sequence exoplanetary atmospheres, these elements do not necessarily reflect the planets' bulk chemical compositions, and planetary atmospheric composition would change in time with white dwarf cooling \citep{kozetal2018}. Other potential future avenues for complementing the chemical constraints from pollution include linking main-sequence stellar composition with planet formation \citep{sanetal2017,hinunt2018,kunetal2018,liuetal2018,cabetal2019} and detecting chemical elements in sublimated tails of closely orbiting disintegrating bodies \citep{vanetal2014,bodetal2018,coletal2018,ridetal2019}.

\subsubsection{Dynamics of white dwarf metal pollution}

The chemical constraints above, in addition to dynamical considerations, support the canonical assumption that the accreted material primarily arises from exo-asteroids as opposed to other classes of extrasolar bodies. 

Several observational and theoretical studies reinforce this conclusion.
An exo-asteroid is currently observed to be disintegrating while orbiting the white dwarf WD 1145+017 at its Roche radius \citep{vanetal2015}, {\rev and other ones are thought to be producing the debris around ZTF J0139+5245 \citep{vanetal2019}.} The dynamical origin of {\rev these bodies are} unknown, but may have been captured by extant extended discs of gas \citep{griver2019}. Additionally, full-lifetime dynamical simulations including the interaction between exo-planets and exo-asteroids \citep{musetal2018} have successfully demonstrated that the latter is accreted onto the white dwarf at a frequency which is commensurate with observations \citep{holetal2018}. Systems with few exo-asteroids can still produce pollution over prolonged timescales through chaotic self-disruption en route to the white dwarf on high-eccentricity orbits, providing intermittent streams of polluting material {\rev \citep{makver2019,veretal2020a}}.

Other types of polluters may occasionally contribute to the metal budget in white dwarf atmospheres. An exo-planet breaking up around a white dwarf {\rev \citep{malper2020a,malper2020b}} and accreting onto the star would showcase a much stronger observational signature than an exo-asteroid which suffers the same fate \citep{beasok2013,beasok2015}. However, {\rev despite the exciting discovery of an ice giant in a compact orbit around a white dwarf \citep{ganetal2019},} exo-planets are not numerous enough for {\rev their destruction} to occur at a high-enough frequency \citep{debsig2002,veretal2013,voyetal2013,musetal2014,vergae2015,veretal2016,veretal2018a} to match the fraction (25-50 per cent) of the observable Milky Way single white dwarfs which feature metal pollution \citep{zucetal2003,zucetal2010,koeetal2014}. 

Exo-moons which experience post-main-sequence evolution can be stripped from their parent exo-planets and then collide with the white dwarf \citep{payetal2016,payetal2017}. However, like exo-planets, exo-moons are not numerous enough (at least if our solar system moons are representative) to explain widespread metal pollution. 

An exo-Oort cloud comet, on average, will impact a white dwarf every $10^4$ yr \citep{alcetal1986,veretal2014a}, providing only a secondary contribution to accreted metals. This statement holds true even if this value varies significantly based on different models and assumptions \citep{paralc1998,stoetal2015,caihey2017}.

\subsubsection{Small body reservoirs}

The reservoirs of bodies smaller than exo-asteroids, however, remain largely unexplored. Some of these reservoirs will carry over from the giant branch phase of stellar evolution {\rev \citep{bonwya2010,maretal2020,veretal2020b}}, where the enhanced stellar luminosity will easily spin up 100~m --- 10~km bodies within 7 au to their breaking point due to the YORP effect {\rev \citep{veretal2014b,versch2020}}. Further, extant debris discs from main sequence evolution have been observed orbiting giant stars at distances of tens of au \citep{bonetal2013,bonetal2014}. 

During the white dwarf phase, boulders, pebbles and dust are subject to radiative drag or Yarkovsky effects that gradually draw them into the star \citep{donetal2010,veretal2015a,veretal2015b,veretal2019a}; upon approach these objects may fragment or sublimate before impacting the photosphere \citep{broetal2017}. Consequently, nearly all pollutant progenitors contribute to the formation and evolution of a white dwarf debris disc, of which over 40 have now been observed {\rev \citep{farihi2016,manetal2020}}.

Gravitational instabilities amongst exo-planets can thrust exo-asteroids towards white dwarfs and instigate periods of bombardments\footnote{The influence of binary stellar companions in post-main-sequence planetary systems can also produce gravitational instabilities, sometimes through the Lidov-Kozai mechanism \citep{kraper2012,vertou2012,musetal2013,portegieszwart2013,bonver2015,hampor2016,kosetal2016,petmun2017,steetal2017,veretal2017a,veretal2017b,steetal2018}.}. Analogously, in the early stages of the solar system, the planetesimal disc was destabilised by the movements of the giant planets \citep{gometal2005}. Lingering uncertainties about the timeline of the Late Heavy Bombardment even within our own solar system {\rev \citep{moretal2018,desetal2020}} indicate that the details of similar events in post-main-sequence exo-systems are even more unconstrained. 

However, the process of bombardment will nevertheless likely occur due to these instabilities. Projectiles will crash into both exo-asteroids and exo-planets, create craters, and eject material back into the interplanetary medium. This material in turn can be radiatively dragged towards the white dwarf or perturbed into it by the extant planetary system architecture. These processes can transport the chemical signature of the target surface to the white dwarf photosphere, where this signature may be measured.

\subsection{Impact ejecta models}

The mechanics of the impact ejecta process in white dwarf planetary systems has not previously been considered in detail. 
In contrast, impact ejecta has been the subject of extensive investigations
within the solar system, as well as a few in other types of extrasolar planetary systems.

\subsubsection{Previous studies}

The physics of impact-induced mass loss from planetary bodies is best constrained from within the solar system. The origin of
the lunar and martian meteorites has provided strong motivation for many investigations 
\citep[e.g.][]{marcus1969,okeahr1977,melosh1984,vicmel1987,heaetal2002,artiva2004,decetal2007,artshu2008,decarli2013,kuretal2018}. Lithopanspermia provides another motivation, both inside and outside of the solar system
\citep{melosh1988,melosh2003,krietal2017,linloe2017,veretal2018b}.

In general, impacts onto target bodies occur obliquely, and the most probable impact angle of randomly incident impactors is 
45 degrees \citep{shoemaker1962}. The amount of ejecta depends on impact angle, as does the structure of the 
resultant crater and the peak shock pressure contour profiles \citep{burmac1998,piemel2000a,greetal2002,schwro2012,takkat2019}. As for the amount of material ejected,
some trends emerge: the ejecta speed decreases as the impact angle increases to $90^{\circ}$, which 
corresponds to a vertical impact (see e.g. Figs. 6-7 of \citealt*{andetal2003}). \cite{artiva2004} found that the
amount of escaping material is minimized for vertical or near-vertical impacts (see their Table 1).

In addition to depending on the impact angle, the amount of mass ejected can be partitioned into three stages depending on ejection timing, location, velocity and experienced shock pressure \citep[e.g.][]{johetal2014,kuretal2018,kurtak2019}: 

\begin{enumerate}

\item Jetting \citep{kieffer1977,melson1986,ang1990,vickery1993,sugsch1999,johetal2014,johetal2015,kuretal2015},  
 
\item Spallation \citep{melosh1984,melosh1985a,vicmel1987,polahr1990,heaetal2002,artiva2004,decetal2007,decarli2013}, and 

\item Normal excavation \citep[e.g.][]{maxwell1977,croft1980,houetal1983,melosh1985b,cinetal1999,andetal2004,houhol2011,tsuetal2015,kurtak2019}. 

\end{enumerate}

\noindent{}Extensive published work about each of these stages provides an indication of the complexity
of impact cratering. 

\subsubsection{Current study}

This level of complexity far exceeds what could be reasonably applied in extrasolar planetary systems, where the 
amount of available data is very sparse compared to that in the solar system. Therefore, we need to make
broad assumptions in order to quantify this idea of generating metal pollution from impact ejecta.

Our three main assumptions are:

\begin{enumerate}

\item We assume vertical impacts on a plane-parallel surface. As previously demonstrated \citep{andetal2003,artiva2004}, doing
so provides a lower bound on the ejecta mass. This bound will usefully help future investigators assess
the plausibility of white dwarf pollution arising from impact ejecta in particular systems.

\item We adopt an analytical model, rather than perform laboratory experiments or run hydrodynamics codes.
An analytical model will allow us to sample a large region of phase space that is relevant to white dwarf
planetary systems, which would otherwise not be possible in a single paper with other methods. 
We use the full-chain analytical model from \citep{kurtak2019}. The advantage of this model is 
that it allows for the high-speed cutoff of the ejecta in the regime of normal excavation to be estimated.

\item We neglect jetting and spallation. Unfortunately, neither simple analytical models for jetting and spallation nor tabular data under a wide range of impact conditions have been established, despite much progress and detailed laboratory and numerical impact experiments \citep[e.g.][]{polahr1990,heaetal2002,artiva2004,artshu2008,kuretal2015,kuretal2018}. Hence, we assume that the large-scale mass ejection is dominated by normal excavation. Now we elaborate on this assumption.

\end{enumerate}

\begin{figure}
\includegraphics[width=8cm]{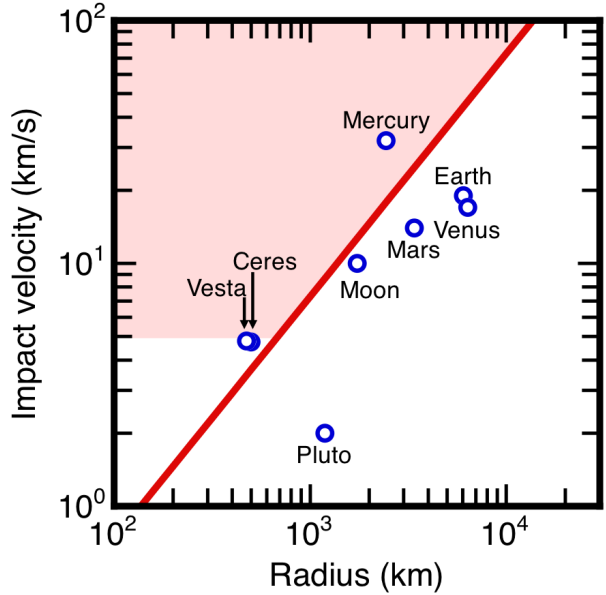}
\caption{
Required impact speed to satisfy the condition that the majority of the total ejecta mass arises from normal excavation. The radius on the $x$-axis is that of the target body ($R_{\rm T}$). The blue circles show the mean impact speed on the bodies in the solar system \citep{itomal2006,obrsyk2011,greetal2015}. The red shaded regions indicate where the results of Kurosawa \& Takada (2019) are applicable to the current study. 
}
\label{ExcPlot}
\end{figure}

In order to develop a criterion for mass ejecta being dominated by normal excavation,
we first need to determine in what parameter regime normal excavation produces the greatest amount of ejecta mass. To do so,
we need to establish the ejecta velocity distribution. Typically, this distribution is given by 
a power-law relation \citep{houetal1983,houhol2011}. 

\cite{kurtak2019} provided an improvement by establishing a high-speed cutoff for this power-law. This upper limit was determined by the residual speed in the isobaric core, which --- near the impact point --- is estimated to be about 10 per cent of the impact speed. Thus, the applicable range of their results in a radius-impact velocity plane can be easily calculated as follows. The required impact speed 
$\left| \vec{v}_{\rm I} \right|$ for the condition that the normal excavation is dominant in the total ejecta mass is given by 

\begin{equation}
\left| \vec{v}_{\rm I} \right| > \frac{v_{\rm esc}}{\alpha}
\label{vIcrit}
\end{equation}

\noindent{}where the escape speed from the target is

\begin{equation}
v_{\rm esc} = \sqrt{\frac{2GM_{\rm T}}{R_{\rm T}}} = R_{\rm T} \sqrt{\frac{8}{3}G\pi \rho_{\rm T}},
\label{escvel}
\end{equation}

\noindent{}such that $\alpha$ is the ratio of the high-speed cutoff to the impact velocity, and $M_{\rm T}$, $R_{\rm T}$ and $\rho_{\rm T}$ are the mass, radius and density of the target. Figure \ref{ExcPlot} illustrates $\left| \vec{v}_{\rm I} \right|$ as a function of $R_{\rm T}$. In this figure, we adopted $\alpha = 0.1$ and $\rho = 3000$ kg/m$^3$. The mean impact speeds for planetary bodies in the solar system are also plotted for context. 

The plot illustrates, as an example, that in our solar system, on average, asteroids in the main belt, as well as Mercury, would meet the condition. The red shaded region illustrates combinations of impact speed and target radii that are applicable in this paper, and will be used to inform the initial conditions of our simulations.

\subsection{Plan for paper}

{\rev In Sections 2 and 3, we respectively outline some additional assumptions and describe the model outputs. Then, in Section 4, we perform computations and summarise our results, primarily through a series of tables. We then address the observational constraints in Section 5, and provide two concrete examples of how this formalism may be used in Section 6.} We discuss the results in {\rev Section 7, and conclude in Section 8}. In order to aid the reader, we have listed descriptions of all of the variables used in this paper in Table \ref{description}.

\begin{table*}
 \centering
 \begin{minipage}{180mm}
  \centering
  \caption{Variables used in this paper.}
  \label{description}
  \begin{tabular}{@{}lll@{}}
  \hline
   Variable & Explanation & Reference \\
 \hline
 $A$ & Auxiliary variable & Equation (\ref{quadA})  \\[2pt]
 $B$ & Auxiliary variable & Equation (\ref{quadB})  \\[2pt]
 $c_{\rm I}$ & Bulk sound speed in the impactor & Table \ref{param}  \\[2pt]
 $c_{\rm T}$ & Bulk sound speed in the target & Table \ref{param}  \\[2pt]
 $C_{\rm T,c}$ & Dimensionless cold coefficient of the target & Table \ref{param}  \\[2pt]
 $C_{\rm T,h}$ & Dimensionless thermal coefficient of the target & Table \ref{param}  \\[2pt]
 $d$ & Distance from the white dwarf at which collision occurs & Chosen by user \\[2pt]
 $D$ & Horizontal distance along target surface to impact point & Chosen by user  \\[2pt]
 $\Delta D$ & Streamtube width along target surface to impact point & Chosen by user  \\[2pt]
 $E_{\rm gra}$ & Gravitational energy & Equation (\ref{Egra})  \\[2pt]
 $E_{\rm kin}$ & Kinetic energy & Equation (\ref{Ekin})  \\[2pt]
 $E_{\rm max}$ & Maximum impact energy to survive target break-up & Equation (\ref{Emax})  \\[2pt]
 $F_{\rm I}$ & Dimensionless coefficient for impactor & Table \ref{param}  \\[2pt]
 $F_{\rm T}$ & Dimensionless coefficient for target & Table \ref{param}  \\[2pt]
 $g_{\rm T}$ & Gravitational acceleration at surface of target & Equation (\ref{surfg})  \\[2pt]
 $K$ & Auxiliary variable & Equation (\ref{quadK})  \\[2pt]
 $l$ & Auxiliary variable & Equation (\ref{auxl})  \\[2pt]
 $m_{\rm T,c}$ & Dimensionless cold exponent of target & Table \ref{param}  \\[2pt]
 $m_{\rm T,h}$ & Dimensionless thermal exponent of target & Table \ref{param}  \\[2pt]
 $M_{\rm exc}$ & Total mass of the excavated (not necessarily escaping) material & Equation (\ref{mexc})   \\[2pt]
 $M_{\rm I}$ & Mass of the impactor & Chosen by user  \\[2pt]
 $M_{\rm tube}$ & Mass in a streamtube & Equation (\ref{Mtube})  \\[2pt]
 $M_{\rm T}$ & Mass of the target & Chosen by user  \\[2pt]
 $n$ & Dimensionless shock decay exponent & Chosen by user  \\[2pt]
 $R_{\rm I}$ & Radius of the impactor & Chosen by user  \\[2pt]
 $R_{\rm T}$ & Radius of the target & Chosen by user  \\[2pt]
 $R_{\rm WD}$ & Radius of the white dwarf & Chosen by user  \\[2pt]
 $u_{\rm T,sw}$ & Switching speed of target & Table \ref{param}  \\[2pt]
 $u_{\rm T,th}$ & Threshold speed of target & Table \ref{param}  \\[2pt]
 $u_{\rm T,0}$ & The peak target material speed in an isobaric core & Equation (\ref{uI})  \\[2pt]
 $\vec{v}_{\rm E}$ & Ejecta velocity & Equation (\ref{ejevel})  \\[2pt]
 $\vec{v}_{\rm I}$ & Impact velocity & Chosen by user  \\[2pt]
 $v_{\rm esc}$ & Escape speed from surface of target & Equation (\ref{escvel})  \\[2pt]
 $v_{\rm orb}$ & Orbital speed of the target & Equation (\ref{maxorb})  \\[2pt]
 $Z$ & Dimensionless streamline shape coefficient & Chosen by user  \\[2pt]
 $\epsilon$ & Auxiliary variable & Equation (\ref{auxep})  \\[2pt]
 $\theta$ & Impactor angle, assuming $0^{\circ}$ is parallel to target surface  & Chosen by user  \\[2pt] 
 $\rho_{\rm T}$ & Density of target  & Table (\ref{param}) \\[2pt]
 $\phi$ & Ejecta angle, assuming $0^{\circ}$ is parallel to target surface & Equation (\ref{phiangle}) \\[2pt] 
\hline
\end{tabular}
\end{minipage}
\end{table*}

\begin{table*}
 \centering
 \begin{minipage}{180mm}
   \centering
   \caption{Physical parameters for four different types of materials.}
   \label{param}
  \begin{tabular}{lcccc}
  \hline
Property & Granite & Basalt & Water (at 25$^{\circ}$C) & Iron \\
 \hline
 $c$ (km/s)          & 3.68 & 2.6  & 2.393 & 3.80 \\[2pt]
 $C_{\rm T,c}$          & 0.0412 & 0.0281 & 0.0796 & 0.0235 \\[2pt]
 $C_{\rm T,h}$          & 0.0833 & 0.0827 & 0.145 & 0.0691 \\[2pt]
 $m_{\rm T,c}$          & 1.97 & 2.29 & 1.39 & 2.14 \\[2pt]
 $m_{\rm T,h}$          & 1.20 & 1.21 & 1.04 & 1.25 \\[2pt]
 $F$                 & 1.24 & 1.62 & 1.333 & 1.58 \\[2pt]
 $u_{\rm T,th}$ (km/s) & 0.804 & 0.730 & 1.48 & 0.652 \\[2pt]
 $u_{\rm T,sw}$ (km/s) & 2.12 & 2.67 & 5.03 & 3.19 \\[2pt]
 $\rho$ (kg/m$^3$)   & 2630 & 2860 & 998 & 7680 \\[2pt]
\hline
\end{tabular}
\end{minipage}
\end{table*}

\section{Additional model assumptions}

We now outline some additional model assumptions.

\subsection{Impactor-to-target radii ratio}

Our model relies on the plane-parallel geometry approximation, which is thought to be valid
for a sufficiently small impactor relative to the size of the target. The approximation holds
when $R_{\rm I}/R_{\rm T}\lesssim 0.2$, where $R_{\rm I}$ and $R_{\rm T}$ are respectively
the radii of the impactor and target \citep{bieetal2013,maretal2014}. We, however, 
take a more conservative approach: we perform computations only for systems
where all of the mass which escapes the target is excavated within $0.2R_{\rm T}$. Throughout 
the manuscript, variables or parameters with subscripts of $I$ and $T$ will refer to properties 
of the impactor and target respectively.

\subsection{Atmosphere-less targets}

We consider only atmosphere-less targets, ranging in size from 1 km up to $10^3$ km (a Pluto-like object).
These objects are typically not expected to have retained a primordial atmosphere at such a late stage of stellar
evolution.

Over the course of a stellar lifetime, stars can strip away atmospheres of planets and moons
which are within a critical distance of the star. This critical distance increases significantly during the
giant branch phase of evolution\footnote{This radiation also changes the chemical
composition of the portions of atmospheres which survive \citep{cametal1988,spimad2012}.}. 
Investigations which have considered this critical distance
{\rev \citep{livsok1984,goldstein1987,neltau1998,soker1998,villiv2007,wicetal2010,beasok2011,schetal2019} }
(see Section 6.1.2. of \citealt*{veras2016b} for a summary) have not converged on a specific expression
for this value. The ability of a planet to retain a certain fraction of its atmosphere throughout
stellar evolution remains an open question. 

Regardless, one may reasonably assume based on the above studies that rocky bodies 
which have survived the giant branch phases of stellar evolution intact within tens of au of their 
parent stars have lost their atmospheres.

\subsection{Impact speed range}

Our choice of impact speed $\left| \vec{v}_{\rm I} \right|$,  which is a relative speed between the impactor
and target, is dictated by equation (\ref{vIcrit}). This speed is weakly dependent on
$d$, the distance from the white dwarf to the collision site. Although increasing $d$ likely helps to
decrease $\left| \vec{v}_{\rm I} \right|$, there is no direct correlation without specification of the
geometry of the collision and the geometry of the orbit. Hence, $d$ does not represent
an input into our simulations.  

Instead the range of $\left| \vec{v}_{\rm I} \right|$ that we provide
spans most realistic values. The lowest value is 5 km/s, which is
the lower limit at which the full-chain model from \cite{kurtak2019}
is applicable. For an upper bound, Eq. 7 of \cite{veretal2014c} illustrates 
that an impactor could
reach speeds relative to the white dwarf of up to 1000 km/s. \footnote{The reason
is because exo-asteroids have to pass inside the Roche radius of the white dwarf
in order to eventually pollute it, and the Roche radius for a typical rubble pile is located
at a distance of about 0.005 au \citep{veretal2017b}.} Our formalism is
not applicable for such high-velocity impacts, but can model collisions
with impact speeds reaching $100$ km/s. 

That value of 100 km/s is still relatively high for white
dwarf planetary systems. For example, the
maximum speed of any orbit occurs at its orbital pericentre, where (as derived by assuming
an orbital eccentricity of unity and neglecting the mass of the target)

\[
{\rm max}\left(v_{\rm orb}\right)  \approx \sqrt{\frac{2 G M_{\rm WD}}{d}}
\]

\begin{equation}
\ \ \ \ \ \ \ \ \ \ \ \ \ \  = 14.6 \ \frac{\rm km}{\rm s} \left( \frac{M_{\rm WD}}{0.6M_{\odot}}  \right)^{1/2}
                                                             \left( \frac{d}{5 \ {\rm au}}  \right)^{-1/2}.
\label{maxorb}
\end{equation}

By providing values for $\left|\vec{v}_{\rm I}\right|$, rather than making any assumptions about
elliptic orbits, we also allow for the possibility
for a collision to occur between one or two bodies on parabolic or hyperbolic
orbits\footnote{One or both bodies could be heading out of the system on a hyperbolic orbit 
following a gravitational instability.}. 


\subsection{Neglecting vaporized mass}

Impactors can melt and vaporize some of the mass which would otherwise escape the target. 
The amount of vaporized mass is a function
of both the impact speed and angle. As demonstrated in \cite{piemel2000b},
the volume of vaporized mass increases with impact angle, peaking at $90^{\circ}$. Other
experiments have established scaling laws as a function of this angle \citep{schultz1996},
also peaking at $90^{\circ}$.

As suggested by \cite{pieetal1997}, vaporization does not become important
until $\left|\vec{v}_{\rm I}\right| \gtrsim 50$ km/s, and only for material which
lies within a couple of $R_{\rm I}$ of the impact point. Therefore, if most of the would-be-ejected
material is close to the impact point, then vaporization would significantly affect 
our high-impact speed computations. 

In fact, the reality is otherwise: as we will show, the amount of ejecta mass which escapes the 
target increases with distance from the impact point, usually for distances of at least tens of $R_{\rm I}$.
Hence, the fraction of escaped mass which would have been vaporized by the impact is typically
negligible. Regardless, in order to be conservative, we report simulation results only for impact events
where mass escapes out to a horizontal distance of at least $15R_{\rm I}$. 

Also, {\rev a key consideration in cratering mechanics is the dissipation of the expanding shock wave
that is produced by the impact. The pressure decay of this wave is parametrized by a 
dimensionless exponent $n$ that will appear in many of the subsequent equations. Rather than
use a constant value for $n$, we instead adopt the following, more realistic speed-dependent exponent from}
\cite{pieetal1997} in our computations:

\begin{equation}
n = -1.84 + 2.61 \log{ \left( \left| \frac{v_{\rm I}}{{\rm km/s}} \right| \right)}
.
\label{npie}
\end{equation}

\noindent{}Further, \cite{pieetal1997} reported that the numerical values in equation (\ref{npie})
are relatively insensitive to the target material.

\section{Model outputs}

Now we discuss in detail some of the properties of the full-chain analytical model from \cite{kurtak2019}
that we adopt here.

\subsection{Ejecta direction}

As in \cite{kurtak2019}, we utilise the Maxwell $Z$-model. {\rev This model represents a way to analytically characterize streamlines in cratering physics, and was borne out of empirical relations noticed in experiments performed by the United States Defense Nuclear Agency \citep{maxwell1977}. The value of $Z$ is mapped to a streamline containing ejecta which} is thrust outward at an angle of 

\begin{equation}
\phi = \tan^{-1}\left(Z-2\right)
.
\label{phiangle}
\end{equation}

\noindent{}Here $\phi = 0^{\circ}$ corresponds to the target surface, and $Z$ 
ranges from 2-5 ($\phi = 0.0^{\circ}-71.6^{\circ}$) {\rev because that range is consistent
with values measured in laboratory experiments \citep{cinetal1999,yametal2017}. Further,}  
$Z = 3$ is adopted for illustration purposes
because it yields the ejection angle of 45 degrees, a typical value observed in {\rev these} 
laboratory experiments.
Note that the angle $\phi$ is independent
of $D$, the distance along the (assumed flat) target surface to the impact point.

\subsection{Ejecta mass}

\noindent{}The mass in a specific streamtube
 $M_{\rm tube}$ between the distances $(D-\Delta D)$ and $D$ is given by

\begin{equation}
M_{\rm tube}(D-\Delta D; D) = 2 \pi \left( \frac{Z-2}{Z+1} \right) \rho_{\rm T} D^2 \Delta D
,
\label{Mtube}
\end{equation}

\noindent{}where $\rho_{\rm T}$ is the density of target and $\Delta D$ is the streamtube width
along the surface. This interval notation will be used throughout
the manuscript for clarity.


Any melted mass will be solidified rapidly after the pressure release, and we neglect vaporized mass as explained above.  The cumulative mass excavated from the surface due to a single impact can then be expressed as $M_{\rm exc}$, where, for a fixed $\Delta D$,

\begin{equation}
M_{\rm exc} = \sum_{D = \Delta D} \left[
M_{\rm tube}(D-\Delta D; D) 
                       \right] 
.
\label{mexc}
\end{equation}

Determining the fraction of $M_{\rm exc}$ which actually escapes the target is a main goal of the paper, and depends on the ejecta speed. For a numerical implementation of equation (\ref{mexc}), we set the upper bound of the summation to $0.2 R_{\rm T}$ (see Section 2.1). One can choose the value of $\Delta D$ depending on the accuracy of the final result which is sought; different values of $\Delta D$ can be easily implemented given the computational speed of the analytical model.

\subsection{Ejecta velocity}

The ejecta velocity ($\vec{v}_{\rm E}$) 
includes direction through the angle $\phi$ given
by equation (\ref{phiangle}) and magnitude through 
Eq. (20) of \cite{kurtak2019}. They provide this magnitude within
a streamtube in terms of the kinetic and gravitational energies 
($E_{\rm kin}$, $E_{\rm gra}$)\footnote{They also
incorporate strength energy, but have caveats about its applicability
within the model. We henceforth neglect the internal strength of our targets.} 
within that same streamtube as

\[
\left| \vec{v}_{\rm E}(D-\Delta D; D) \right| = 
\]

\begin{equation}
\ \ \ 
\sqrt{
\frac{2\left[E_{\rm kin}(D-\Delta D; D) - E_{\rm gra}(D-\Delta D; D)\right]}
{M_{\rm tube}(D-\Delta D; D)}
}
.
\label{ejevel}
\end{equation}

These individual energies are given by Eqs. (27-36) of \cite{kurtak2019}, and each include
the arbitrary lengthscale $\Delta D$ which cancels out in the computation of 
$\left|\vec{v}_{\rm E}(D-\Delta D; D)\right|$ for a particular streamtube.
Therefore, $\left|\vec{v}_{\rm E}(D-\Delta D; D)\right|$ is independent of $\Delta D$, but is dependent on $D$,
and tends towards 0 as $D$ increases. 

The gravitational energy 

\begin{equation}
E_{\rm gra}(D-\Delta D; D) = \pi \left[\frac{Z^2-4Z+4}{Z\left(Z+2\right)}\right] g_{\rm T} \rho_{\rm T} D^3 \Delta D
\label{Egra}
\end{equation}

\noindent{}is a function of the density $\rho_{\rm T}$ and surface gravity $g_{\rm T}$ of the target, the latter of
which is given by

\begin{equation}
g_{\rm T} = \frac{GM_{\rm T}}{R_{\rm T}^2}
,
\label{surfg}
\end{equation}

\noindent{}with $M_{\rm T}$ representing the mass of the target. Further, all targets are assumed to be spheres.
This assumption allows one to easily compute the target mass ($M_{\rm T}$) given $\rho_{\rm T}$ and $R_{\rm T}$.

The kinetic energy ($E_{\rm kin}$) contains several terms, and these in turn include several dimensionless
constants and speeds. They all refer to properties of the target ($m_{\rm T,c}, m_{\rm T,h}, C_{\rm T,c}, C_{\rm T,h}$), where
the subscripts ``c'' and  ``h'' refer to ``cold'' and ``hot''.
Two of the speeds ($u_{\rm T,sw}, u_{\rm T,th}$:  ``switching'' and ``threshold'') also refer to properties of the target material. All these values are given for a variety of materials (granite, basalt, water 
and iron) in Table \ref{param} and described in physical detail in \cite{kurtak2019}.

One important speed is $u_{\rm T,0}$, which is the peak target material speed in an isobaric core. This speed is dependent on several variables, including the impact speed $\left|\vec{v}_{\rm I}\right|$ (technically, the speed of the impactor at the target surface). The value of $u_{\rm T,0}$ is computed by using the formulation on Pg. 231 of \cite{melosh2011} (originally from \citealt*{melosh1989}), which illustrates

\[
u_{\rm T,0} = \frac{-B + \sqrt{B^2 - 4AK}}{2A}, \ \ \ {\rm if} \ A\ne 0
\]

\begin{equation}
\ \ \ \ \ \ \, = \frac{\left|\vec{v}_{\rm I} \right|}{2}, \ \ \ \ \ \ \ \ \ \ \ \ \ \ \ \ \ 
            \ \ \ \ \ \ \ {\rm if} \ A=0
\label{uI}
\end{equation}

\noindent{}where

\begin{equation}
A \equiv \rho_{\rm T} F_{\rm T} - \rho_{\rm I} F_{\rm I}
,
\label{quadA}
\end{equation}

\begin{equation}
B \equiv \rho_{\rm T} c_{\rm T} + \rho_{\rm I} c_{\rm I} + 2 \rho_{\rm I} F_{\rm I} \left|\vec{v}_{\rm I} \right|
,
\label{quadB}  
\end{equation}

\begin{equation}
K \equiv -\rho_{\rm I} \left|\vec{v}_{\rm I} \right| \left(c_{\rm I} + F_{\rm I} \left|\vec{v}_{\rm I} \right|  \right)
.
\label{quadK}
\end{equation}

\noindent{}The values $c_{\rm T}, c_{\rm I}$, and $F_{\rm I}$ represent more parameters specific to the target and the impactor, and are given in Table \ref{param} (reproduced from \citealt*{melosh2011}). Note that for equal-composition impactors and targets, $A=0$.

The kinetic energy is either, for $u_{\rm T,0} > u_{\rm T, sw}$,

\[
E_{\rm kin}(D-\Delta D; D) = R_{\rm I}^{Z+1} D^{1-Z} \Delta D \bigg\lbrace 
\pi \left[\frac{Z-2}{Z+1}\right] \rho_{\rm T} C_{\rm T,h}^2 u_{\rm T,0}^{2m_{\rm T,h}} 
\]

\[
+ 
\left[\frac{\pi\left(Z-2\right)\rho_{\rm T} C_{\rm T,h}^2  }{2 n m_{\rm T,h}-Z-1}\right]  
\left( 
u_{\rm T,0}^{2m_{\rm T,h}}
+
u_{\rm T,0}^{\frac{Z+1}{n}} u_{\rm T,sw}^{\frac{2nm_{\rm T,h}-Z-1}{n}}
\right)
\]

\[
+ 
\left[\frac{\pi\left(Z-2\right)\rho_{\rm T} C_{\rm T,c}^2   }{2 n m_{\rm T,c}-Z-1}\right]  \left(u_{\rm T,sw}^{\frac{2nm_{\rm T,c}-Z-1}{n}} - u_{\rm T,th}^{\frac{2nm_{\rm T,c}-Z-1}{n}} \right) u_{\rm T,0}^{\frac{Z+1}{n}}
\bigg\rbrace
,
\]

\begin{equation}
\label{Ekin}
\end{equation}

\noindent{}or, for $u_{\rm T,0} < u_{\rm T, sw}$,

\

\[
E_{\rm kin}(D-\Delta D; D) = \pi R_{\rm I}^{Z+1} D^{1-Z} \Delta D \rho_{\rm T}C_{\rm T,c}^2
\bigg\lbrace 
\left[\frac{Z-2}{Z+1}\right] u_{\rm T,0}^{2m_{\rm T,c}}
\]

\[
+
\left[\frac{Z-2}{2 n m_{\rm T,c}-Z-1}\right]  
\left( 
u_{\rm T,0}^{2m_{\rm T,c}}
+
u_{\rm T,0}^{\frac{Z+1}{n}} u_{\rm T,th}^{\frac{2nm_{\rm T,c}-Z-1}{n}}
\right)
\bigg\rbrace
.
\]

\begin{equation}
\label{Ekin2}
\end{equation}

\subsection{Escaping the target}

Excavated material will escape the gravitational pull of the target only when $\left|\vec{v}_{\rm E} \right| > v_{\rm esc}$.
Because $\left| \vec{v}_{\rm E} \right|$ is a function of $D$, there will be a critical value of $D$ beyond which escape is not possible, where the ejecta will re-impact the target. Computing this critical value of $D$ allows us to determine the cumulative escaped mass from all inward streamtubes. {\rev If mass
does escape the target, then it will escape with an ``excess speed'' equal to 
$\sqrt{\left|v_{\rm E}\right|^2 - v_{\rm esc}^2}$ for a given streamline.} 

\section{Simulation inputs and results}

By using the formalism of the last section, we can compute the amount of mass ejected
due to a single impact. Doing so first involves solving equation (\ref{ejevel}) for each 
streamtube. We set 
$\Delta D = 10^{-4} \times \left(0.2 R_{\rm T}\right)$ uniformly across all simulations, 
which achieves a balance of accuracy and speed across the wide range of parameter space
studied. We emphasise that in order to obtain a higher accuracy for any particular setup, the computation
should be rerun with progressively smaller $\Delta D$ until the desired result is acquired.

\subsection{Input variables}

Only in rare cases will the direction
of the escaped material be towards the white dwarf, which represents an Earth-sized target.
Instead, the ejected matter will populate reservoirs of debris that may themselves be perturbed at a later
time towards the white dwarf, or slowly dragged into the white dwarf due to Poynting-Robertson drag.
Hence, when modelling debris trajectories -- and keeping in mind that vertical impacts yield a lower bound
for the amount of escaped mass -- one must couple the escape direction
with the orbital architecture, the geometry of the collision, and the excess speed.
We consider escaped mass values across the entire range $Z=2.01-5$ at seven values, thereby
avoiding the singularities which occur at $Z=2$.





\begin{table*}
 \centering
 \begin{minipage}{180mm}
   \centering
   \caption{Mass which escapes into the interplanetary medium after a collision between a spherical basaltic impactor and a spherical basaltic target, where $R_{\rm I} = 10^{-5}R_{\rm T} $, for given target radii, impact speeds, and the entire range of $Z$ (streamtube shape) values. Results are given in {\rev both kg and (immediately below in parenthesis)} the impactor mass $M_{\rm I}$. ``APART'' refers to cases where the target would be destroyed,  ``LIMIT'' where the plane-parallel approximation of the full-chain analytical model breaks down, ``JET'' where the impact speed fails to satisfy equation (\ref{vIcrit}), and ``VAPOR'' for where the amount of vaporized mass may not be negligible. See the main text for more details.
}
   \label{tabfid10m5}
  \begin{tabular}{lccccccc}
\hline
  &   \multicolumn{7}{c}{Cumulative escaped mass in terms {\rev of both kg and $M_{\rm I}$}}    \\[2pt]
\hline
\hline
 $\mathbf{ R_{\rm I} = 10^{-5}R_{\rm T} }$   & $Z=2.01$ & $Z=2.5$ & $Z=3.0$ & $Z=3.5$ & $Z=4.0$ & $Z=4.5$ & $Z=5.0$  \\[2pt]
\hline
 $R_{\rm T} = 1$ km  &   \multicolumn{7}{c}{ }    \\[2pt]
\hline 
\ \ \ $\left|\vec{v}_{\rm I}\right| = 5$ km/s     &  $3.2 \times 10^7$ kg  &  $2.8 \times 10^7$ kg &  $2.3 \times 10^6$ kg      
&   $3.0 \times 10^5$ kg  &  $5.4 \times 10^4$ kg  &   $1.3 \times 10^4$ kg &  $4.1 \times 10^3$ kg  \\[2pt]
    \ \ \ & $(2.7 \times 10^9 M_{\rm I})$   & $(2.3 \times 10^9 M_{\rm I})$ & $(1.9 \times 10^8 M_{\rm I})$ 
& $(2.5 \times 10^7 M_{\rm I})$ & $(4.5 \times 10^6 M_{\rm I})$ &  $(1.1 \times 10^6 M_{\rm I})$ & $(3.4 \times 10^5 M_{\rm I})$ \\[2pt]

\ \ \ $\left|\vec{v}_{\rm I}\right| = 10$ km/s   &   LIMIT  &  $1.8 \times 10^{9}$ kg  &  $1.9 \times 10^{8}$ kg                         
&   $3.0 \times 10^7$ kg  &  $6.6 \times 10^6$ kg &   $1.9 \times 10^6$ kg  &  $6.7 \times 10^5$ kg  \\[2pt]
\ \ \ & LIMIT   & $(1.5 \times 10^{11} M_{\rm I})$ & $(1.6 \times 10^{10} M_{\rm I})$ 
& $(2.5 \times 10^9 M_{\rm I})$ & $(5.5 \times 10^{8} M_{\rm I})$ &  $(1.6 \times 10^8 M_{\rm I})$ & $(5.6 \times 10^7 M_{\rm I})$ \\[2pt]

\ \ \ $\left|\vec{v}_{\rm I}\right| = 30$ km/s   &   LIMIT  &  $3.0 \times 10^{8}$ kg  &  $3.0 \times 10^7$ kg                        
&   $4.4 \times 10^6$ kg &   $9.0 \times 10^5$ kg  &   $2.4 \times 10^5$ kg &  $7.9 \times 10^4$ kg  \\[2pt]
\ \ \ & LIMIT   & $(2.5 \times 10^{10} M_{\rm I})$ & $(2.5 \times 10^{9} M_{\rm I})$ 
& $(3.7 \times 10^8 M_{\rm I})$ & $(7.5 \times 10^{7} M_{\rm I})$ &  $(2.0 \times 10^7 M_{\rm I})$ & $(6.6 \times 10^6 M_{\rm I})$ \\[2pt]

\ \ \ $\left|\vec{v}_{\rm I}\right| = 60$ km/s   &   LIMIT  &  $2.0 \times 10^{8}$ kg  &  $2.2 \times 10^7$ kg                           
&   $3.1 \times 10^6$ kg &  $6.5 \times 10^5$ kg  &   $1.7 \times 10^5$ kg  &  $5.8 \times 10^4$ kg  \\[2pt]
\ \ \ & LIMIT   & $(1.7 \times 10^{10} M_{\rm I})$ & $(1.8 \times 10^{9} M_{\rm I})$ 
& $(2.6 \times 10^8 M_{\rm I})$ & $(5.4 \times 10^{7} M_{\rm I})$ &  $(1.4 \times 10^7 M_{\rm I})$ & $(4.8 \times 10^6 M_{\rm I})$ \\[2pt]

\ \ \ $\left|\vec{v}_{\rm I}\right| = 100$ km/s  &  $1.4 \times 10^{8}$ kg  &  $1.7 \times 10^{8}$ kg  &  $1.8 \times 10^7$ kg            
& $2.6 \times 10^6$ kg  &   $5.5 \times 10^5$ kg   &   $1.4 \times 10^5$ kg &  $5.0 \times 10^4$ kg \\[2pt]
\ \ \ & $(1.2 \times 10^{10} M_{\rm I})$   & $(1.4 \times 10^{10} M_{\rm I})$ & $(1.5 \times 10^{9} M_{\rm I})$ 
& $(2.2 \times 10^8 M_{\rm I})$ & $(4.6 \times 10^{7} M_{\rm I})$ &  $(1.2 \times 10^7 M_{\rm I})$ & $(4.2 \times 10^6 M_{\rm I})$ \\[2pt]

\hline
 $R_{\rm T} = 10$ km  &   \multicolumn{7}{c}{ }    \\[2pt]
\hline
\ \ \ $\left|\vec{v}_{\rm I}\right| = 5$ km/s     &  $3.4 \times 10^8$ kg   &  $5.3 \times 10^8$ kg  &  $7.4 \times 10^7$ kg                         
&   $1.4 \times 10^7$ kg  &  $3.5 \times 10^6$ kg  &   $1.1 \times 10^6$ kg &  $4.0 \times 10^5$ kg  \\[2pt]
\ \ \ & $(2.8 \times 10^{7} M_{\rm I})$   & $(4.4 \times 10^{7} M_{\rm I})$ & $(6.2 \times 10^{6} M_{\rm I})$ 
& $(1.2 \times 10^6 M_{\rm I})$ & $(2.9 \times 10^{5} M_{\rm I})$ &  $(8.9 \times 10^4 M_{\rm I})$ & $(3.3 \times 10^4 M_{\rm I})$ \\[2pt]

\ \ \ $\left|\vec{v}_{\rm I}\right| = 10$ km/s   &   $1.8 \times 10^{10}$ kg  &  $3.5 \times 10^{10}$ kg  &  $6.1 \times 10^9$ kg                         
&   $1.4 \times 10^9$ kg &  $4.2 \times 10^8$ kg  &   $1.6 \times 10^8$ kg  &  $6.6 \times 10^7$ kg \\[2pt]
\ \ \ & $(1.5 \times 10^{9} M_{\rm I})$   & $(2.9 \times 10^{9} M_{\rm I})$ & $(5.1 \times 10^{8} M_{\rm I})$ 
& $(1.2 \times 10^8 M_{\rm I})$ & $(3.5 \times 10^{7} M_{\rm I})$ &  $(1.3 \times 10^7 M_{\rm I})$ & $(5.5 \times 10^6 M_{\rm I})$ \\[2pt]

\ \ \ $\left|\vec{v}_{\rm I}\right| = 30$ km/s   &   $2.8 \times 10^9$ kg  &  $5.8 \times 10^9$ kg  &  $9.6 \times 10^8$ kg                       
&   $2.0 \times 10^8$ kg &   $5.6 \times 10^7$ kg &   $1.9 \times 10^7$ kg  &  $7.7 \times 10^6$ kg  \\[2pt]
\ \ \ & $(2.3 \times 10^{8} M_{\rm I})$   & $(4.8 \times 10^{8} M_{\rm I})$ & $(8.0 \times 10^{7} M_{\rm I})$ 
& $(1.7 \times 10^7 M_{\rm I})$ & $(4.7 \times 10^{6} M_{\rm I})$ &  $(1.6 \times 10^6 M_{\rm I})$ & $(6.4 \times 10^5 M_{\rm I})$ \\[2pt]

\ \ \ $\left|\vec{v}_{\rm I}\right| = 60$ km/s   &   $1.8 \times 10^9$ kg  &  $3.8 \times 10^9$ kg  &  $6.7 \times 10^8$ kg                          
&   $1.4 \times 10^8$ kg &  $4.1 \times 10^7$ kg   &   $1.4 \times 10^7$ kg &  $5.9 \times 10^6$ kg \\[2pt]
\ \ \ & $(1.5 \times 10^{8} M_{\rm I})$   & $(3.2 \times 10^{8} M_{\rm I})$ & $(5.6 \times 10^{7} M_{\rm I})$ 
& $(1.2 \times 10^7 M_{\rm I})$ & $(3.4 \times 10^{6} M_{\rm I})$ &  $(1.2 \times 10^6 M_{\rm I})$ & $(4.9 \times 10^5 M_{\rm I})$ \\[2pt]

\ \ \ $\left|\vec{v}_{\rm I}\right| = 100$ km/s  &  $1.4 \times 10^9$ kg &  $3.2 \times 10^9$ kg &  $5.5 \times 10^8$ kg                       
& $1.2 \times 10^8$ kg  &   $3.5 \times 10^7$ kg  &   $1.2 \times 10^7$ kg  &  $5.0 \times 10^6$ kg \\[2pt]
\ \ \ & $(1.2 \times 10^{8} M_{\rm I})$   & $(2.7 \times 10^{8} M_{\rm I})$ & $(4.6 \times 10^{7} M_{\rm I})$ 
& $(1.0 \times 10^7 M_{\rm I})$ & $(2.9 \times 10^{6} M_{\rm I})$ &  $(1.0 \times 10^6 M_{\rm I})$ & $(4.2 \times 10^5 M_{\rm I})$ \\[2pt]

\hline
 $R_{\rm T} = 100$ km  &   \multicolumn{7}{c}{ }    \\[2pt]
\hline
\ \ \ $\left|\vec{v}_{\rm I}\right| = 5$ km/s     &  $3.4 \times 10^9$ kg  &  $1.0 \times 10^{10}$ kg &  $2.3 \times 10^9$ kg                          
&   $6.5 \times 10^8$ kg  &  $2.2 \times 10^8$ kg &   $8.1 \times 10^7$ kg &  $3.6 \times 10^7$ kg  \\[2pt]
\ \ \ & $(2.8 \times 10^{5} M_{\rm I})$   & $(8.6 \times 10^{5} M_{\rm I})$ & $(1.9 \times 10^{5} M_{\rm I})$ 
& $(5.4 \times 10^4 M_{\rm I})$ & $(1.8 \times 10^{4} M_{\rm I})$ &  $(6800 M_{\rm I})$ & $(3000 M_{\rm I})$ \\[2pt]

\ \ \ $\left|\vec{v}_{\rm I}\right| = 10$ km/s   &   $1.8 \times 10^{11}$ kg  &  $6.8 \times 10^{11}$ kg &  $1.9 \times 10^{11}$ kg                         
&   $6.5 \times 10^{10}$ kg  &  $2.6 \times 10^{10}$ kg &   $1.2 \times 10^{10}$ kg  &  $6.7 \times 10^9$ kg  \\[2pt]
\ \ \ & $(1.5 \times 10^{7} M_{\rm I})$   & $(5.7 \times 10^{7} M_{\rm I})$ & $(1.6 \times 10^{7} M_{\rm I})$ 
& $(5.4 \times 10^6 M_{\rm I})$ & $(2.2 \times 10^{6} M_{\rm I})$ &  $(1.0 \times 10^6 M_{\rm I})$ & $(5.6 \times 10^5 M_{\rm I})$ \\[2pt]

\ \ \ $\left|\vec{v}_{\rm I}\right| = 30$ km/s   &   $2.8 \times 10^{10}$ kg &  $1.1 \times 10^{11}$ kg  &  $3.0 \times 10^{10}$ kg                        
&   $9.6 \times 10^9$ kg &   $3.6 \times 10^9$ kg  &   $1.6 \times 10^9$ kg  &  $7.4 \times 10^8$ kg \\[2pt]
\ \ \ & $(2.3 \times 10^{6} M_{\rm I})$   & $(9.2 \times 10^{6} M_{\rm I})$ & $(2.5 \times 10^{6} M_{\rm I})$ 
& $(8.0 \times 10^5 M_{\rm I})$ & $(3.0 \times 10^{5} M_{\rm I})$ &  $(1.3 \times 10^5 M_{\rm I})$ & $(6.2 \times 10^4 M_{\rm I})$ \\[2pt]

\ \ \ $\left|\vec{v}_{\rm I}\right| = 60$ km/s   &   $1.8 \times 10^{10}$ kg  &  $7.5 \times 10^{10}$ kg  &  $2.2 \times 10^{10}$ kg                           
&   $6.8 \times 10^9$ kg &  $2.6 \times 10^9$ kg  &   $1.2 \times 10^9$ kg  &  $5.5 \times 10^8$ kg \\[2pt]
\ \ \ & $(1.5 \times 10^{6} M_{\rm I})$   & $(6.3 \times 10^{6} M_{\rm I})$ & $(1.8 \times 10^{6} M_{\rm I})$ 
& $(5.7 \times 10^5 M_{\rm I})$ & $(2.2 \times 10^{5} M_{\rm I})$ &  $(9.6 \times 10^4 M_{\rm I})$ & $(4.6 \times 10^4 M_{\rm I})$ \\[2pt]

\ \ \ $\left|\vec{v}_{\rm I}\right| = 100$ km/s  &  $1.4 \times 10^{10}$ kg  &  $6.1 \times 10^{10}$ kg  &  $1.8 \times 10^{10}$ kg                        
& $5.9 \times 10^9$ kg  &   $2.2 \times 10^9$ kg   &   $9.7 \times 10^8$ kg  &  $5.0 \times 10^8$ kg  \\[2pt]
\ \ \ & $(1.2 \times 10^{6} M_{\rm I})$   & $(5.1 \times 10^{6} M_{\rm I})$ & $(1.5 \times 10^{6} M_{\rm I})$ 
& $(4.9 \times 10^5 M_{\rm I})$ & $(1.8 \times 10^{5} M_{\rm I})$ &  $(8.1 \times 10^4 M_{\rm I})$ & $(4.2 \times 10^4 M_{\rm I})$ \\[2pt]

\hline
 $R_{\rm T} = 10^3$ km  &   \multicolumn{7}{c}{ }    \\[2pt]
\hline
\ \ \ $\left|\vec{v}_{\rm I}\right| = 5$ km/s     &  JET   & JET  &   JET                         
&    JET  &  JET  &   JET  &  JET   \\[2pt]
\ \ \ &  JET   & JET  &   JET                         
&    JET  &  JET  &   JET  &  JET   \\[2pt]

\ \ \ $\left|\vec{v}_{\rm I}\right| = 10$ km/s   &   $1.8 \times 10^{12}$ kg  &  $1.3 \times 10^{13}$ kg &  $6.0 \times 10^{12}$ kg                         
&   $3.0 \times 10^{12}$ kg  &  $1.7 \times 10^{12}$ kg &   $9.7 \times 10^{11}$ kg  &  $6.8 \times 10^{11}$ kg   \\[2pt]
\ \ \ & $(1.5 \times 10^{5} M_{\rm I})$   & $(1.1 \times 10^{6} M_{\rm I})$ & $(5.0 \times 10^{5} M_{\rm I})$ 
& $(2.5 \times 10^5 M_{\rm I})$ & $(1.4 \times 10^{5} M_{\rm I})$ &  $(8.1 \times 10^4 M_{\rm I})$ & $(5.7 \times 10^4 M_{\rm I})$ \\[2pt]

\ \ \ $\left|\vec{v}_{\rm I}\right| = 30$ km/s   &   $2.8 \times 10^{11}$ kg  &  $2.2 \times 10^{12}$ kg  &  $9.8 \times 10^{11}$ kg                        
&   $4.6 \times 10^{11}$ kg &   $2.2 \times 10^{11}$ kg &   $1.4 \times 10^{11}$ kg  &  $7.3 \times 10^{10}$ kg \\[2pt]
\ \ \ & $(2.3 \times 10^{4} M_{\rm I})$   & $(1.8 \times 10^{5} M_{\rm I})$ & $(8.2 \times 10^{4} M_{\rm I})$ 
& $(3.8 \times 10^4 M_{\rm I})$ & $(1.8 \times 10^{4} M_{\rm I})$ &  $(1.2 \times 10^4 M_{\rm I})$ & $(6100 M_{\rm I})$ \\[2pt]

\ \ \ $\left|\vec{v}_{\rm I}\right| = 60$ km/s   &   $1.9 \times 10^{11}$ kg &  $1.4 \times 10^{12}$ kg &  $6.8 \times 10^{11}$ kg                           
&   $3.4 \times 10^{11}$ kg &  $1.7 \times 10^{11}$ kg   &   $9.8 \times 10^{10}$ kg &  $5.9 \times 10^{10}$ kg \\[2pt]
\ \ \ & $(1.6 \times 10^{4} M_{\rm I})$   & $(1.2 \times 10^{5} M_{\rm I})$ & $(5.7 \times 10^{4} M_{\rm I})$ 
& $(2.8 \times 10^4 M_{\rm I})$ & $(1.4 \times 10^{4} M_{\rm I})$ &  $(8200 M_{\rm I})$ & $(4900 M_{\rm I})$ \\[2pt]

\ \ \ $\left|\vec{v}_{\rm I}\right| = 100$ km/s  &  $1.4 \times 10^{11}$ kg  &  $1.2 \times 10^{12}$ kg  &  $5.9 \times 10^{11}$ kg                       
& $2.6 \times 10^{11}$ kg  &   $1.4 \times 10^{11}$ kg   &   $8.1 \times 10^{10}$ kg  &  $4.7 \times 10^{10}$ kg \\[2pt]
\ \ \ & $(1.2 \times 10^{4} M_{\rm I})$   & $(9.8 \times 10^{4} M_{\rm I})$ & $(4.9 \times 10^{4} M_{\rm I})$ 
& $(2.2 \times 10^4 M_{\rm I})$ & $(1.2 \times 10^{4} M_{\rm I})$ &  $(6800 M_{\rm I})$ & $(3900 M_{\rm I})$ \\[2pt]

\hline
\end{tabular}
\end{minipage}
\end{table*}

\begin{table*}
 \centering
 \begin{minipage}{180mm}
   \centering
   \caption{
Same as Table \ref{tabfid10m5}, but for $R_{\rm I} = 10^{-4}R_{\rm T}$.
}
   \label{tabfid10m4}
  \begin{tabular}{lccccccc}
\hline
  &   \multicolumn{7}{c}{Cumulative escaped mass in terms {\rev of both kg and $M_{\rm I}$}}    \\[2pt]
\hline
\hline
 $\mathbf{ R_{\rm I} = 10^{-4}R_{\rm T} }$   & $Z=2.01$ & $Z=2.5$ & $Z=3.0$ & $Z=3.5$ & $Z=4.0$ & $Z=4.5$ & $Z=5.0$  \\[2pt]
\hline
 $R_{\rm T} = 1$ km  &   \multicolumn{7}{c}{ }    \\[2pt]
\hline 
\ \ \ $\left|\vec{v}_{\rm I}\right| = 5$ km/s     &  LIMIT   &  LIMIT  &  $2.3 \times 10^9$ kg                         
&   $3.0 \times 10^8$ kg  &  $5.4 \times 10^7$ kg  &   $1.3 \times 10^7$ kg &  $4.1 \times 10^6$ kg  \\[2pt]
\ \ \ &  LIMIT   &  LIMIT & $(1.9 \times 10^{8} M_{\rm I})$ 
& $(2.5 \times 10^7 M_{\rm I})$ & $(4.5 \times 10^{6} M_{\rm I})$ &  $(1.1 \times 10^6 M_{\rm I})$ & $(3.4 \times 10^5 M_{\rm I})$ \\[2pt]

\ \ \ $\left|\vec{v}_{\rm I}\right| = 10$ km/s   &   LIMIT   &  LIMIT   &   LIMIT                         
&   LIMIT   &  $6.5 \times 10^9$ kg  &   $1.9 \times 10^9$ kg  &  $6.6 \times 10^8$ kg  \\[2pt]
\ \ \ &  LIMIT   &  LIMIT & LIMIT 
& LIMIT & $(5.4 \times 10^{8} M_{\rm I})$ &  $(1.6 \times 10^8 M_{\rm I})$ & $(5.5 \times 10^7 M_{\rm I})$ \\[2pt]

\ \ \ $\left|\vec{v}_{\rm I}\right| = 30$ km/s   &   LIMIT  &  LIMIT  &    LIMIT                      
&   $4.3 \times 10^9$ kg &   $8.9 \times 10^8$ kg  &   $2.4 \times 10^8$ kg &  $7.8 \times 10^7$ kg \\[2pt]
\ \ \ &  LIMIT   &  LIMIT & LIMIT 
& $(3.6 \times 10^8 M_{\rm I})$ & $(7.4 \times 10^{7} M_{\rm I})$ &  $(2.0 \times 10^7 M_{\rm I})$ & $(6.5 \times 10^6 M_{\rm I})$ \\[2pt]

\ \ \ $\left|\vec{v}_{\rm I}\right| = 60$ km/s   &   LIMIT  &   LIMIT &   LIMIT                         
&   $3.1 \times 10^9$ kg &  $6.5 \times 10^8$ kg  &   $1.7 \times 10^8$ kg  &  $5.8 \times 10^7$ kg  \\[2pt]
\ \ \ &  LIMIT   &  LIMIT & LIMIT 
& $(2.6 \times 10^8 M_{\rm I})$ & $(5.4 \times 10^{7} M_{\rm I})$ &  $(1.4 \times 10^7 M_{\rm I})$ & $(4.8 \times 10^6 M_{\rm I})$ \\[2pt]

\ \ \ $\left|\vec{v}_{\rm I}\right| = 100$ km/s  &  LIMIT  &  LIMIT  &  LIMIT                        
& $2.6 \times 10^9$ kg &   $5.5 \times 10^8$ kg   &   $1.4 \times 10^8$ kg  &  $4.9 \times 10^7$ kg \\[2pt]
\ \ \ &  LIMIT   &  LIMIT & LIMIT 
& $(2.2 \times 10^8 M_{\rm I})$ & $(4.6 \times 10^{7} M_{\rm I})$ &  $(1.2 \times 10^7 M_{\rm I})$ & $(4.1 \times 10^6 M_{\rm I})$ \\[2pt]

\hline
 $R_{\rm T} = 10$ km  &   \multicolumn{7}{c}{ }    \\[2pt]
\hline
\ \ \ $\left|\vec{v}_{\rm I}\right| = 5$ km/s     &  LIMIT   &  $5.3 \times 10^{11}$ kg &  $7.3 \times 10^{10}$ kg                          
&   $1.3 \times 10^{10}$ kg  &  $3.4 \times 10^9$ kg  &   $1.1 \times 10^9$ kg &  $4.0 \times 10^8$ kg   \\[2pt]
\ \ \ &  LIMIT   &  $(4.4 \times 10^{7} M_{\rm I})$ & $(6.1 \times 10^{6} M_{\rm I})$ 
& $(1.1 \times 10^6 M_{\rm I})$ & $(2.8 \times 10^{5} M_{\rm I})$ &  $(9.0 \times 10^4 M_{\rm I})$ & $(3.3 \times 10^4 M_{\rm I})$ \\[2pt]

\ \ \ $\left|\vec{v}_{\rm I}\right| = 10$ km/s   &   LIMIT  &  LIMIT  &  $6.0 \times 10^{12}$ kg                        
&   $1.3 \times 10^{12}$ kg  &  $4.1 \times 10^{11}$ kg &   $1.6 \times 10^{11}$ kg  &  $6.7 \times 10^{10}$ kg  \\[2pt]
\ \ \ &  LIMIT   &  LIMIT & $(5.0 \times 10^{8} M_{\rm I})$ 
& $(1.1 \times 10^8 M_{\rm I})$ & $(3.4 \times 10^{7} M_{\rm I})$ &  $(1.3 \times 10^7 M_{\rm I})$ & $(5.6 \times 10^6 M_{\rm I})$ \\[2pt]

\ \ \ $\left|\vec{v}_{\rm I}\right| = 30$ km/s   &   LIMIT  &  $5.6 \times 10^{12}$ kg &  $9.6 \times 10^{11}$ kg                        
&   $2.0 \times 10^{11}$ kg &   $5.6 \times 10^{10}$ kg &   $1.9 \times 10^{10}$ kg &  $7.8 \times 10^9$ kg \\[2pt]
\ \ \ &  LIMIT   &  $(4.7 \times 10^{8} M_{\rm I})$ & $(8.0 \times 10^{7} M_{\rm I})$ 
& $(1.7 \times 10^7 M_{\rm I})$ & $(4.7 \times 10^{6} M_{\rm I})$ &  $(1.6 \times 10^6 M_{\rm I})$ & $(6.5 \times 10^5 M_{\rm I})$ \\[2pt]

\ \ \ $\left|\vec{v}_{\rm I}\right| = 60$ km/s   &   LIMIT  &  $3.8 \times 10^{12}$ kg &  $6.7 \times 10^{11}$ kg                           
&   $1.4 \times 10^{11}$ kg &  $4.1 \times 10^{10}$ kg  &   $1.4 \times 10^{10}$ kg  &  $5.8 \times 10^9$ kg  \\[2pt]
\ \ \ &  LIMIT   &  $(3.2 \times 10^{8} M_{\rm I})$ & $(5.6 \times 10^{7} M_{\rm I})$ 
& $(1.2 \times 10^7 M_{\rm I})$ & $(3.4 \times 10^{6} M_{\rm I})$ &  $(1.2 \times 10^6 M_{\rm I})$ & $(4.8 \times 10^5 M_{\rm I})$ \\[2pt]

\ \ \ $\left|\vec{v}_{\rm I}\right| = 100$ km/s  &   LIMIT &  $3.1 \times 10^{12}$ kg  &  $5.5 \times 10^{11}$ kg                        
& $1.2 \times 10^{11}$ kg  &   $3.5 \times 10^{10}$ kg   &   $1.2 \times 10^{10}$ kg  &  $4.9 \times 10^9$ kg \\[2pt]
\ \ \ &  LIMIT   &  $(2.6 \times 10^{8} M_{\rm I})$ & $(4.6 \times 10^{7} M_{\rm I})$ 
& $(1.0 \times 10^7 M_{\rm I})$ & $(2.9 \times 10^{6} M_{\rm I})$ &  $(1.0 \times 10^6 M_{\rm I})$ & $(4.1 \times 10^5 M_{\rm I})$ \\[2pt]

\hline
 $R_{\rm T} = 100$ km  &   \multicolumn{7}{c}{ }    \\[2pt]
\hline

\ \ \ $\left|\vec{v}_{\rm I}\right| = 5$ km/s     &  $3.4 \times 10^{12}$ kg   &  $1.0 \times 10^{13}$ kg &  $2.3 \times 10^{12}$ kg                         
&   $6.3 \times 10^{11}$ kg  &  $2.2 \times 10^{11}$ kg  &   $8.6 \times 10^{10}$ kg  &  $4.0 \times 10^{10}$ kg  \\[2pt]
\ \ \ &  $(2.8 \times 10^5 M_{\rm I})$  &  $(8.5 \times 10^{5} M_{\rm I})$ & $(1.9 \times 10^{5} M_{\rm I})$ 
& $(5.3 \times 10^4 M_{\rm I})$ & $(1.8 \times 10^{4} M_{\rm I})$ &  $(7200 M_{\rm I})$ & $(3300 M_{\rm I})$ \\[2pt]

\ \ \ $\left|\vec{v}_{\rm I}\right| = 10$ km/s   &   LIMIT  &  $6.8 \times 10^{14}$ kg  &  $1.9 \times 10^{14}$ kg                        
&   $6.5 \times 10^{13}$ kg  &  $2.6 \times 10^{13}$ kg  &   $1.2 \times 10^{13}$ kg &  $6.7 \times 10^{12}$ kg  \\[2pt]
\ \ \ &  LIMIT   &  $(5.7 \times 10^{7} M_{\rm I})$ & $(1.6 \times 10^{7} M_{\rm I})$ 
& $(5.4 \times 10^6 M_{\rm I})$ & $(2.2 \times 10^{6} M_{\rm I})$ &  $(1.0 \times 10^6 M_{\rm I})$ & $(5.6 \times 10^5 M_{\rm I})$ \\[2pt]

\ \ \ $\left|\vec{v}_{\rm I}\right| = 30$ km/s   &   $2.9 \times 10^{13}$ kg  &  $1.1 \times 10^{14}$ kg  &  $3.0 \times 10^{13}$ kg                        
&   $9.5 \times 10^{12}$ kg &   $3.6 \times 10^{12}$ kg  &   $1.6 \times 10^{12}$ kg  &  $7.8 \times 10^{11}$ kg \\[2pt]
\ \ \ &  $(2.4 \times 10^6 M_{\rm I})$  &  $(9.2 \times 10^{6} M_{\rm I})$ & $(2.5 \times 10^{6} M_{\rm I})$ 
& $(7.9 \times 10^5 M_{\rm I})$ & $(3.0 \times 10^{5} M_{\rm I})$ &  $(1.3 \times 10^5 M_{\rm I})$ & $(6.5 \times 10^4 M_{\rm I})$ \\[2pt]

\ \ \ $\left|\vec{v}_{\rm I}\right| = 60$ km/s   &   $1.8 \times 10^{13}$ kg  &  $7.5 \times 10^{13}$ kg &  $2.2 \times 10^{13}$ kg                           
&   $6.8 \times 10^{12}$ kg &  $2.6 \times 10^{12}$ kg  &   $1.1 \times 10^{12}$ kg  &  $5.8 \times 10^{11}$ kg  \\[2pt]
\ \ \ &  $(1.5 \times 10^6 M_{\rm I})$  &  $(6.3 \times 10^{6} M_{\rm I})$ & $(1.8 \times 10^{6} M_{\rm I})$ 
& $(5.7 \times 10^5 M_{\rm I})$ & $(2.2 \times 10^{5} M_{\rm I})$ &  $(9.5 \times 10^4 M_{\rm I})$ & $(4.8 \times 10^4 M_{\rm I})$ \\[2pt]

\ \ \ $\left|\vec{v}_{\rm I}\right| = 100$ km/s  &  $1.4 \times 10^{13}$ kg &  $6.1 \times 10^{13}$ kg &  $1.8 \times 10^{13}$ kg                        
& $5.8 \times 10^{12}$ kg  &   $2.2 \times 10^{12}$ kg   &   $9.8 \times 10^{11}$ kg  &  $4.9 \times 10^{11}$ kg \\[2pt]
\ \ \ &  $(1.2 \times 10^6 M_{\rm I})$  &  $(5.1 \times 10^{6} M_{\rm I})$ & $(1.5 \times 10^{6} M_{\rm I})$ 
& $(4.8 \times 10^5 M_{\rm I})$ & $(1.8 \times 10^{5} M_{\rm I})$ &  $(8.2 \times 10^4 M_{\rm I})$ & $(4.1 \times 10^4 M_{\rm I})$ \\[2pt]

\hline
 $R_{\rm T} = 10^3$ km  &   \multicolumn{7}{c}{ }    \\[2pt]
\hline
\ \ \ $\left|\vec{v}_{\rm I}\right| = 5$ km/s     &  JET   & JET  &   JET                         
&    JET &  JET  &   JET  &  JET   \\[2pt]
\ \ \ &  JET   & JET  &   JET                         
&    JET &  JET  &   JET  &  JET   \\[2pt]

\ \ \ $\left|\vec{v}_{\rm I}\right| = 10$ km/s   &   $1.8 \times 10^{15}$ kg  &  $1.3 \times 10^{16}$ kg  &  $6.1 \times 10^{15}$ kg                         
&   $3.0 \times 10^{15}$ kg &  $1.7 \times 10^{15}$ kg  &   $1.0 \times 10^{15}$ kg  &  $6.7 \times 10^{14}$ kg   \\[2pt]
\ \ \ &  $(1.5 \times 10^5 M_{\rm I})$  &  $(1.1 \times 10^{6} M_{\rm I})$ & $(5.1 \times 10^{5} M_{\rm I})$ 
& $(2.5 \times 10^5 M_{\rm I})$ & $(1.4 \times 10^{5} M_{\rm I})$ &  $(8.4 \times 10^4 M_{\rm I})$ & $(5.6 \times 10^4 M_{\rm I})$ \\[2pt]

\ \ \ $\left|\vec{v}_{\rm I}\right| = 30$ km/s   &   $2.9 \times 10^{14}$ kg  &  $2.2 \times 10^{15}$ kg  &  $9.6 \times 10^{14}$ kg                        
&   $4.4 \times 10^{14}$ kg &   $2.3 \times 10^{14}$ kg  &   $1.3 \times 10^{14}$ kg  &  $7.8 \times 10^{13}$ kg  \\[2pt]
\ \ \ &  $(2.4 \times 10^4 M_{\rm I})$  &  $(1.8 \times 10^{5} M_{\rm I})$ & $(8.0 \times 10^{4} M_{\rm I})$ 
& $(3.7 \times 10^4 M_{\rm I})$ & $(1.9 \times 10^{4} M_{\rm I})$ &  $(1.1 \times 10^4 M_{\rm I})$ & $(6500 M_{\rm I})$ \\[2pt]

\ \ \ $\left|\vec{v}_{\rm I}\right| = 60$ km/s   &   $1.9 \times 10^{14}$ kg &  $1.4 \times 10^{15}$ kg  &  $6.7 \times 10^{14}$ kg                           
&   $3.1 \times 10^{14}$ kg &  $1.7 \times 10^{14}$ kg   &   $9.3 \times 10^{13}$ kg &  $5.8 \times 10^{13}$ kg  \\[2pt]
\ \ \ &  $(1.6 \times 10^4 M_{\rm I})$  &  $(1.2 \times 10^{5} M_{\rm I})$ & $(5.6 \times 10^{4} M_{\rm I})$ 
& $(2.6 \times 10^4 M_{\rm I})$ & $(1.4 \times 10^{4} M_{\rm I})$ &  $(7800 M_{\rm I})$ & $(4800 M_{\rm I})$ \\[2pt]

\ \ \ $\left|\vec{v}_{\rm I}\right| = 100$ km/s  &  $1.4 \times 10^{14}$ kg  &  $1.2 \times 10^{15}$ kg  &  $5.5 \times 10^{14}$ kg                        
& $2.6 \times 10^{14}$ kg  &   $1.4 \times 10^{14}$ kg   &   $8.0 \times 10^{13}$ kg &  $4.9 \times 10^{13}$ kg \\[2pt]
\ \ \ &  $(1.2 \times 10^4 M_{\rm I})$  &  $(9.9 \times 10^{4} M_{\rm I})$ & $(4.6 \times 10^{4} M_{\rm I})$ 
& $(2.2 \times 10^4 M_{\rm I})$ & $(1.2 \times 10^{4} M_{\rm I})$ &  $(6700 M_{\rm I})$ & $(4100 M_{\rm I})$ \\[2pt]

\hline
& & & & & & &\\[2pt]
& & & & & & &\\[2pt]
& & & & & & &\\[2pt]
& & & & & & &\\[2pt]
& & & & & & &\\[2pt]
\end{tabular}
\end{minipage}
\end{table*}

\begin{table*}
 \centering
 \begin{minipage}{180mm}
   \centering
   \caption{Same as Table \ref{tabfid10m4}, but for $R_{\rm I} = 10^{-3}R_{\rm T}$.}
   \label{tabfid10m3}
  \begin{tabular}{lccccccc}
\hline
    &   \multicolumn{7}{c}{Cumulative escaped mass in terms {\rev of both kg and $M_{\rm I}$}}    \\[2pt]
\hline
\hline
$\mathbf{ R_{\rm I} = 10^{-3}R_{\rm T} }$    & $Z=2.01$ & $Z=2.5$ & $Z=3.0$ & $Z=3.5$ & $Z=4.0$ & $Z=4.5$ & $Z=5.0$  \\[2pt]
\hline
 $R_{\rm T} = 1$ km  &   \multicolumn{7}{c}{ }    \\[2pt]
\hline
\ \ \ $\left|\vec{v}_{\rm I}\right| = 5$ km/s     &  LIMIT   &   LIMIT  &    LIMIT                        
&   LIMIT   &   LIMIT  &   $1.3 \times 10^{10}$ kg &  $4.0 \times 10^9$ kg   \\[2pt]
\ \ \ &  LIMIT   &   LIMIT  &    LIMIT                        
&   LIMIT   &   LIMIT  &   $(1.1 \times 10^6 M_{\rm I})$  &  $(3.3 \times 10^5 M_{\rm I})$   \\[2pt]

\ \ \ $\left|\vec{v}_{\rm I}\right| = 10$ km/s   &   LIMIT   &  LIMIT   &    LIMIT                       
&    LIMIT  &   LIMIT  &   LIMIT   &  LIMIT    \\[2pt]
\ \ \ &   LIMIT   &  LIMIT   &    LIMIT                       
&    LIMIT  &   LIMIT  &   LIMIT   &  LIMIT    \\[2pt]

\ \ \ $\left|\vec{v}_{\rm I}\right| = 30$ km/s   &    LIMIT  &   LIMIT  &   LIMIT                        
&    LIMIT &    LIMIT &    LIMIT &   LIMIT \\[2pt]
\ \ \ &   LIMIT   &  LIMIT   &    LIMIT                       
&    LIMIT  &   LIMIT  &   LIMIT   &  LIMIT    \\[2pt]

\ \ \ $\left|\vec{v}_{\rm I}\right| = 60$ km/s   &     LIMIT &   LIMIT &    LIMIT                         
&   LIMIT &    LIMIT &     LIMIT &   LIMIT \\[2pt]
\ \ \ &   LIMIT   &  LIMIT   &    LIMIT                       
&    LIMIT  &   LIMIT  &   LIMIT   &  LIMIT    \\[2pt]

\ \ \ $\left|\vec{v}_{\rm I}\right| = 100$ km/s  &  APART  &  APART  &  APART                        
& APART  &   APART   &   APART  &  APART \\[2pt]
\ \ \ &  APART  &  APART  &  APART                        
& APART  &   APART   &   APART  &  APART \\[2pt]

\hline
 $R_{\rm T} = 10$ km  &   \multicolumn{7}{c}{ }    \\[2pt]
\hline
\ \ \ $\left|\vec{v}_{\rm I}\right| = 5$ km/s     &   LIMIT  &   LIMIT &   LIMIT                         
&   $1.3 \times 10^{13}$ kg &  $3.4 \times 10^{12}$ kg  &   $1.1 \times 10^{12}$ kg &  $4.0 \times 10^{11}$ kg   \\[2pt]
\ \ \ &  LIMIT   &   LIMIT  &    LIMIT                        
&   $(1.1 \times 10^6 M_{\rm I})$    &   $(2.8 \times 10^5 M_{\rm I})$   &   $(8.8 \times 10^4 M_{\rm I})$  &  $(3.3 \times 10^4 M_{\rm I})$   \\[2pt]

\ \ \ $\left|\vec{v}_{\rm I}\right| = 10$ km/s   &    LIMIT &   LIMIT &     LIMIT                     
&    LIMIT &   LIMIT &    LIMIT &   LIMIT  \\[2pt]
\ \ \ &    LIMIT &   LIMIT &     LIMIT                     
&    LIMIT &   LIMIT &    LIMIT &   LIMIT  \\[2pt]

\ \ \ $\left|\vec{v}_{\rm I}\right| = 30$ km/s   &    LIMIT &   LIMIT &      LIMIT                    
&    LIMIT &   LIMIT &   $1.9 \times 10^{13}$ kg &  $7.7 \times 10^{12}$ kg  \\[2pt]
\ \ \ &  LIMIT   &   LIMIT  &    LIMIT                        
&   LIMIT    &   LIMIT   &   $(1.6 \times 10^6 M_{\rm I})$  &  $(6.4 \times 10^5 M_{\rm I})$   \\[2pt]

\ \ \ $\left|\vec{v}_{\rm I}\right| = 60$ km/s   &   LIMIT  &  LIMIT &     LIMIT                        
&    LIMIT &   LIMIT  &   $1.4 \times 10^{13}$ kg &  $5.6 \times 10^{12}$ kg \\[2pt]
\ \ \ &  LIMIT   &   LIMIT  &    LIMIT                        
&   LIMIT    &   LIMIT   &   $(1.2 \times 10^6 M_{\rm I})$  &  $(4.7 \times 10^5 M_{\rm I})$   \\[2pt]

\ \ \ $\left|\vec{v}_{\rm I}\right| = 100$ km/s  &  LIMIT  &  LIMIT &    LIMIT                      
&   LIMIT &    LIMIT  &   $1.2 \times 10^{13}$ kg  &  $4.9 \times 10^{12}$ kg \\[2pt]
\ \ \ &  LIMIT   &   LIMIT  &    LIMIT                        
&   LIMIT    &   LIMIT   &   $(9.9 \times 10^5 M_{\rm I})$  &  $(4.1 \times 10^5 M_{\rm I})$   \\[2pt]

\hline
 $R_{\rm T} = 100$ km  &   \multicolumn{7}{c}{ }    \\[2pt]
\hline
\ \ \ $\left|\vec{v}_{\rm I}\right| = 5$ km/s     &   LIMIT   &  LIMIT  &  $2.3 \times 10^{15}$ kg                          
&   $6.3 \times 10^{14}$ kg &  $2.2 \times 10^{14}$ kg  &   $8.6 \times 10^{13}$ kg  &  $4.0 \times 10^{13}$ kg   \\[2pt]
\ \ \ &  LIMIT   &   LIMIT  &    $(1.9 \times 10^5 M_{\rm I})$                        
&   $(5.3 \times 10^4 M_{\rm I})$    &   $(1.8 \times 10^4 M_{\rm I})$   &   $(7200 M_{\rm I})$  &  $(3300 M_{\rm I})$   \\[2pt]

\ \ \ $\left|\vec{v}_{\rm I}\right| = 10$ km/s   &   LIMIT  &   LIMIT &     LIMIT                     
&  LIMIT  &  LIMIT  &   $1.2 \times 10^{16}$ kg &  $6.6 \times 10^{15}$ kg   \\[2pt]
\ \ \ &  LIMIT   &   LIMIT  &    LIMIT                        
&   LIMIT    &   LIMIT   &   $(1.0 \times 10^6 M_{\rm I})$  &  $(5.5 \times 10^5 M_{\rm I})$   \\[2pt]

\ \ \ $\left|\vec{v}_{\rm I}\right| = 30$ km/s   &  LIMIT   &  LIMIT  &    LIMIT                     
&   $9.3 \times 10^{15}$ kg &   $3.5 \times 10^{15}$ kg &   $1.6 \times 10^{15}$ kg &  $7.7 \times 10^{14}$ kg  \\[2pt]
\ \ \ &  LIMIT   &   LIMIT  &    LIMIT                        
&   $(7.8 \times 10^5 M_{\rm I})$    &   $(2.9 \times 10^5 M_{\rm I})$   &   $(1.3 \times 10^5 M_{\rm I})$  &  $(6.4 \times 10^4 M_{\rm I})$   \\[2pt]

\ \ \ $\left|\vec{v}_{\rm I}\right| = 60$ km/s   &    LIMIT &  LIMIT  &    LIMIT                         
&   $6.7 \times 10^{15}$ kg &  $2.5 \times 10^{15}$ kg  &   $1.1 \times 10^{15}$ kg &  $5.6 \times 10^{14}$ kg \\[2pt]
\ \ \ &  LIMIT   &   LIMIT  &    LIMIT                        
&   $(5.6 \times 10^5 M_{\rm I})$    &   $(2.1 \times 10^5 M_{\rm I})$   &   $(9.5 \times 10^4 M_{\rm I})$  &  $(4.7 \times 10^4 M_{\rm I})$   \\[2pt]

\ \ \ $\left|\vec{v}_{\rm I}\right| = 100$ km/s  &   LIMIT &   LIMIT &     LIMIT                     
&  $5.6 \times 10^{15}$ kg &   $2.2 \times 10^{15}$ kg   &   $9.7 \times 10^{14}$ kg  &  $4.9 \times 10^{14}$ kg \\[2pt]
\ \ \ &  LIMIT   &   LIMIT  &    LIMIT                        
&   $(4.7 \times 10^5 M_{\rm I})$    &   $(1.8 \times 10^5 M_{\rm I})$   &   $(8.1 \times 10^4 M_{\rm I})$  &  $(4.1 \times 10^4 M_{\rm I})$   \\[2pt]

\hline
 $R_{\rm T} = 10^3$ km  &   \multicolumn{7}{c}{ }    \\[2pt]
\hline
\ \ \ $\left|\vec{v}_{\rm I}\right| = 5$ km/s     &  JET   &  JET  &   JET                         
&    JET  &   JET   &   JET  &   JET   \\[2pt]
\ \ \ &  JET   &  JET  &   JET                         
&    JET  &   JET   &   JET  &   JET \\[2pt]

\ \ \ $\left|\vec{v}_{\rm I}\right| = 10$ km/s   &   LIMIT  &   LIMIT  &  $6.0 \times 10^{18}$ kg                        
&   $3.0 \times 10^{18}$ kg  &  $1.7 \times 10^{18}$ kg  &   $9.9 \times 10^{17}$ kg  &  $6.6 \times 10^{17}$ kg   \\[2pt]
\ \ \ &  LIMIT   &   LIMIT  &    $(5.0 \times 10^5 M_{\rm I})$                       
&   $(2.5 \times 10^5 M_{\rm I})$    &   $(1.4 \times 10^5 M_{\rm I})$   &   $(8.3 \times 10^4 M_{\rm I})$  &  $(5.5 \times 10^4 M_{\rm I})$   \\[2pt]

\ \ \ $\left|\vec{v}_{\rm I}\right| = 30$ km/s   &   LIMIT  &  $2.2 \times 10^{18}$ kg  &  $9.6 \times 10^{17}$ kg                        
&   $4.4 \times 10^{17}$ kg &   $2.3 \times 10^{17}$ kg  &   $1.3 \times 10^{17}$ kg  &  $7.8 \times 10^{16}$ kg  \\[2pt]
\ \ \ &  LIMIT   &   $(1.8 \times 10^5 M_{\rm I})$  &    $(8.0 \times 10^4 M_{\rm I})$                       
&   $(3.7 \times 10^4 M_{\rm I})$    &   $(1.9 \times 10^4 M_{\rm I})$   &   $(1.1 \times 10^4 M_{\rm I})$  &  $(6500 M_{\rm I})$   \\[2pt]

\ \ \ $\left|\vec{v}_{\rm I}\right| = 60$ km/s   &   LIMIT  &  $1.4 \times 10^{18}$ kg &  $6.7 \times 10^{17}$ kg                          
&   $3.1 \times 10^{17}$ kg &  $1.7 \times 10^{17}$ kg  &   $9.2 \times 10^{16}$ kg  &  $5.8 \times 10^{16}$ kg  \\[2pt]
\ \ \ &  LIMIT   &   $(1.2 \times 10^5 M_{\rm I})$  &    $(5.6 \times 10^4 M_{\rm I})$                       
&   $(2.6 \times 10^4 M_{\rm I})$    &   $(1.4 \times 10^4 M_{\rm I})$   &   $(7700 M_{\rm I})$  &  $(4800 M_{\rm I})$   \\[2pt]

\ \ \ $\left|\vec{v}_{\rm I}\right| = 100$ km/s  &  $1.4 \times 10^{17}$ kg &  $1.2 \times 10^{18}$ kg  &  $5.5 \times 10^{17}$ kg                        
& $2.6 \times 10^{17}$ kg  &   $1.4 \times 10^{17}$ kg   &   $7.9 \times 10^{16}$ kg  &  $4.9 \times 10^{16}$ kg \\[2pt]
\ \ \ &  $(1.2 \times 10^4 M_{\rm I})$   &   $(9.8 \times 10^4 M_{\rm I})$  &    $(4.6 \times 10^4 M_{\rm I})$                       
&   $(2.2 \times 10^4 M_{\rm I})$    &   $(1.2 \times 10^4 M_{\rm I})$   &   $(6600 M_{\rm I})$  &  $(4100 M_{\rm I})$   \\[2pt]

\hline
& & & & & & &\\[2pt]
& & & & & & &\\[2pt]
& & & & & & &\\[2pt]
& & & & & & &\\[2pt]
& & & & & & &\\[2pt]
\end{tabular}
\end{minipage}
\end{table*}

\begin{table*}
 \centering
 \begin{minipage}{180mm}
   \centering
   \caption{Escaped post-collisional mass for impacts between a spherical basaltic impactor and a spherical 
target of different materials. See the Table 3 caption for explanations of the non-numeric entries.
In all cases, $R_{\rm T} = 500$~km and $R_{\rm I} = 10^{-3}R_{\rm T} = 0.5$ km.}
   \label{combos}
  \begin{tabular}{lccccccc}
\hline
    &   \multicolumn{7}{c}{Cumulative escaped mass in terms {\rev of both kg and $M_{\rm I}$}}    \\[2pt]
\hline
\hline
    & $Z=2.01$ & $Z=2.5$ & $Z=3.0$ & $Z=3.5$ & $Z=4.0$ & $Z=4.5$ & $Z=5.0$  \\[2pt]
\hline
 Basaltic target  &   \multicolumn{7}{c}{ }    \\[2pt]
\hline
\ \ \ $\left|\vec{v}_{\rm I}\right| = 5$ km/s      &   $1.6 \times 10^{16}$ kg    &  $8.1 \times 10^{16}$ kg  &  $2.5 \times 10^{16}$ kg   &  $9.3 \times 10^{15}$ kg &  $3.9 \times 10^{15}$ kg  &  $1.8 \times 10^{15}$ kg  &  $1.0 \times 10^{15}$ kg      \\[2pt]
\ \ \ &  $(1.1 \times 10^4 M_{\rm I})$   &   $(5.4 \times 10^4 M_{\rm I})$  &    $(1.7 \times 10^4 M_{\rm I})$                       
&   $(6200 M_{\rm I})$    &   $(2600 M_{\rm I})$   &   $(1200 M_{\rm I})$  &  $(670 M_{\rm I})$   \\[2pt]

\ \ \ $\left|\vec{v}_{\rm I}\right| = 10$ km/s     &  LIMIT   &  LIMIT    &   LIMIT   &  $9.3 \times 10^{17}$ kg  &  $4.6 \times 10^{17}$ kg  &  $2.7 \times 10^{17}$ kg  &  $1.6 \times 10^{17}$ kg    \\[2pt]
\ \ \ &  LIMIT   &   LIMIT  &    LIMIT                       
&   $(6.2 \times 10^5 M_{\rm I})$    &   $(3.1 \times 10^5 M_{\rm I})$   &   $(1.8 \times 10^5 M_{\rm I})$  &  $(1.1 \times 10^5 M_{\rm I})$   \\[2pt]

\ \ \ $\left|\vec{v}_{\rm I}\right| = 30$ km/s     &    LIMIT  &  LIMIT  &  $3.3 \times 10^{17}$ kg   &  $1.4 \times 10^{17}$ kg &  $6.3 \times 10^{16}$ kg  &  $3.3 \times 10^{16}$ kg  &  $1.9 \times 10^{16}$ kg   \\[2pt]
\ \ \ &  LIMIT   &   LIMIT  &    $(2.2 \times 10^5 M_{\rm I})$                     
&   $(9.2 \times 10^4 M_{\rm I})$    &   $(4.2 \times 10^4 M_{\rm I})$   &   $(2.2 \times 10^4 M_{\rm I})$  &  $(1.3 \times 10^4 M_{\rm I})$   \\[2pt]

\ \ \ $\left|\vec{v}_{\rm I}\right| = 60$ km/s     &  LIMIT   &   $5.8 \times 10^{17}$ kg  &  $2.4 \times 10^{17}$ kg  &  $9.9 \times 10^{16}$ kg  &  $4.6 \times 10^{16}$ kg  &  $2.4 \times 10^{16}$ kg &  $1.4 \times 10^{16}$ kg     \\[2pt]
\ \ \ &  LIMIT   &   $(3.9 \times 10^5 M_{\rm I})$  &    $(1.6 \times 10^5 M_{\rm I})$                       
&   $(6.6 \times 10^4 M_{\rm I})$    &   $(3.1 \times 10^4 M_{\rm I})$   &   $(1.6 \times 10^4 M_{\rm I})$  &  $(9500 M_{\rm I})$   \\[2pt]

\ \ \ $\left|\vec{v}_{\rm I}\right| = 100$ km/s   &  LIMIT   &  $4.8 \times 10^{17}$ kg  &  $1.9 \times 10^{17}$ kg   &  $8.4 \times 10^{16}$ kg &  $3.9 \times 10^{16}$ kg   &  $2.1 \times 10^{16}$ kg &  $1.2 \times 10^{16}$ kg   \\[2pt]
\ \ \ &  LIMIT   &   $(3.2 \times 10^5 M_{\rm I})$  &    $(1.3 \times 10^5 M_{\rm I})$                       
&   $(5.6 \times 10^4 M_{\rm I})$    &   $(2.6 \times 10^4 M_{\rm I})$   &   $(1.4 \times 10^4 M_{\rm I})$  &  $(8200 M_{\rm I})$   \\[2pt]

\hline
 Granite target  &   \multicolumn{7}{c}{ }    \\[2pt]
\hline
\ \ \ $\left|\vec{v}_{\rm I}\right| = 5$ km/s      &   VAPOR    &  VAPOR  &  VAPOR   &  VAPOR  &  VAPOR  &  VAPOR  &  VAPOR      \\[2pt]
\ \ \ &   VAPOR    &  VAPOR  &  VAPOR   &  VAPOR  &  VAPOR  &  VAPOR  &  VAPOR      \\[2pt]

\ \ \ $\left|\vec{v}_{\rm I}\right| = 10$ km/s     &  $1.6 \times 10^{16}$ kg   &  $1.9 \times 10^{17}$ kg   &  $1.2 \times 10^{17}$ kg   &  $7.5 \times 10^{16}$ kg  &  $4.9 \times 10^{16}$ kg  &  $3.4 \times 10^{16}$ kg  &  $2.5 \times 10^{16}$ kg    \\[2pt]
\ \ \ &  $(1.1 \times 10^4 M_{\rm I})$   &   $(1.3 \times 10^5 M_{\rm I})$  &    $(8.1 \times 10^4 M_{\rm I})$                       
&   $(5.0 \times 10^4 M_{\rm I})$    &   $(3.3 \times 10^4 M_{\rm I})$   &   $(2.3 \times 10^4 M_{\rm I})$  &  $(1.7 \times 10^4 M_{\rm I})$   \\[2pt]

\ \ \ $\left|\vec{v}_{\rm I}\right| = 30$ km/s     &   $1.5 \times 10^{15}$ kg  &  $1.8 \times 10^{16}$ kg  &  $1.2 \times 10^{16}$ kg  &  $7.3 \times 10^{15}$ kg  &  $4.8 \times 10^{15}$ kg  &  $3.3 \times 10^{15}$ kg  &  $2.4 \times 10^{15}$ kg    \\[2pt]
\ \ \ &  $(980 M_{\rm I})$   &   $(1.2 \times 10^4 M_{\rm I})$  &    $(7900 M_{\rm I})$                       
&   $(4900 M_{\rm I})$    &   $(3200 M_{\rm I})$   &   $(2200 M_{\rm I})$  &  $(1600 M_{\rm I})$   \\[2pt]

\ \ \ $\left|\vec{v}_{\rm I}\right| = 60$ km/s      &  $8.5 \times 10^{14}$ kg   &   $1.1 \times 10^{16}$ kg  &  $7.3 \times 10^{15}$ kg  &  $4.6 \times 10^{15}$ kg  &  $3.0 \times 10^{15}$ kg  &  $2.1 \times 10^{15}$ kg  &  $1.5 \times 10^{15}$ kg    \\[2pt]
\ \ \ &  $(570 M_{\rm I})$   &   $(7300 M_{\rm I})$  &    $(4900 M_{\rm I})$                       
&   $(3100 M_{\rm I})$    &   $(2000 M_{\rm I})$   &   $(1400 M_{\rm I})$  &  $(1000 M_{\rm I})$   \\[2pt]

\ \ \ $\left|\vec{v}_{\rm I}\right| = 100$ km/s   &  $6.6 \times 10^{14}$ kg   &  $8.5 \times 10^{15}$ kg  &  $5.8 \times 10^{15}$ kg   &  $3.7 \times 10^{15}$ kg  &  $2.5 \times 10^{15}$ kg   &  $1.6 \times 10^{15}$ kg &  $1.3 \times 10^{15}$ kg   \\[2pt]
\ \ \ &  $(440 M_{\rm I})$   &   $(5700 M_{\rm I})$  &    $(3900 M_{\rm I})$                       
&   $(2500 M_{\rm I})$    &   $(1700 M_{\rm I})$   &   $(1100 M_{\rm I})$  &  $(840 M_{\rm I})$   \\[2pt]

\hline
 Waterworld target  &   \multicolumn{7}{c}{ }    \\[2pt]
\hline
\ \ \ $\left|\vec{v}_{\rm I}\right| = 5$ km/s      &   $2.5 \times 10^{11}$ kg    &  $4.9 \times 10^{12}$ kg  &  $4.6 \times 10^{12}$ kg   &  VAPOR  &  VAPOR  &  VAPOR  &  VAPOR      \\[2pt]
\ \ \ &  $(0.17 M_{\rm I})$   &   $(3.3 M_{\rm I})$  &    $(3.1 M_{\rm I})$                       
&   VAPOR   &   VAPOR   &   VAPOR  &  VAPOR   \\[2pt]

\ \ \ $\left|\vec{v}_{\rm I}\right| = 10$ km/s     &  $2.4 \times 10^{13}$ kg &  $4.9 \times 10^{14}$ kg   &  $5.1 \times 10^{14}$ kg  &  $4.5 \times 10^{14}$ kg &  $3.9 \times 10^{14}$ kg &  $3.4 \times 10^{14}$ kg &  $3.1 \times 10^{14}$ kg   \\[2pt]
\ \ \ &  $(16 M_{\rm I})$   &   $(330 M_{\rm I})$  &    $(340 M_{\rm I})$                       
&   $(300 M_{\rm I})$    &   $(260 M_{\rm I})$   &   $(230 M_{\rm I})$  &  $(210 M_{\rm I})$   \\[2pt]

\ \ \ $\left|\vec{v}_{\rm I}\right| = 30$ km/s     &   $5.8 \times 10^{12}$ kg  &  $1.2 \times 10^{14}$ kg  &  $1.3 \times 10^{14}$ kg  &  $9.0 \times 10^{13}$ kg  &  $8.1 \times 10^{13}$ kg  &  $6.9 \times 10^{13}$ kg  &  $6.0 \times 10^{13}$ kg    \\[2pt]
\ \ \ &  $(3.9 M_{\rm I})$   &   $(82 M_{\rm I})$  &    $(89 M_{\rm I})$                       
&   $(60 M_{\rm I})$    &   $(54 M_{\rm I})$   &   $(46 M_{\rm I})$  &  $(40 M_{\rm I})$   \\[2pt]

\ \ \ $\left|\vec{v}_{\rm I}\right| = 60$ km/s      &  $4.8 \times 10^{12}$ kg &   $9.6 \times 10^{13}$ kg  &  $9.4 \times 10^{13}$ kg  &  $8.1 \times 10^{13}$ kg  &  $6.9 \times 10^{13}$ kg  &  $6.1 \times 10^{13}$ kg  &  $4.0 \times 10^{13}$ kg     \\[2pt]
\ \ \ &  $(3.2 M_{\rm I})$   &   $(64 M_{\rm I})$  &    $(63 M_{\rm I})$                       
&   $(54 M_{\rm I})$    &   $(46 M_{\rm I})$   &   $(41 M_{\rm I})$  &  $(27 M_{\rm I})$   \\[2pt]

\ \ \ $\left|\vec{v}_{\rm I}\right| = 100$ km/s   &  $5.4 \times 10^{12}$ kg   &  $9.9 \times 10^{13}$ kg  &  $9.1 \times 10^{13}$ kg   &  $7.5 \times 10^{13}$ kg  &  $6.1 \times 10^{13}$ kg  &  $5.2 \times 10^{13}$ kg &  $4.5 \times 10^{13}$ kg   \\[2pt]
\ \ \ &  $(3.6 M_{\rm I})$   &   $(66 M_{\rm I})$  &    $(61 M_{\rm I})$                       
&   $(50 M_{\rm I})$    &   $(41 M_{\rm I})$   &   $(35 M_{\rm I})$  &  $(30 M_{\rm I})$   \\[2pt]

\hline
 Iron target  &   \multicolumn{7}{c}{ }    \\[2pt]
\hline

\ \ \ $\left|\vec{v}_{\rm I}\right| = 5$ km/s      &   $1.8 \times 10^{14}$ kg   &  $1.8 \times 10^{15}$ kg  &  $1.1 \times 10^{15}$ kg  &  $6.0 \times 10^{14}$ kg  &  $3.7 \times 10^{14}$ kg  &  $2.4 \times 10^{14}$ kg  &  VAPOR      \\[2pt]
\ \ \ &  $(118 M_{\rm I})$   &   $(1200 M_{\rm I})$  &    $(710 M_{\rm I})$                       
&   $(400 M_{\rm I})$    &   $(250 M_{\rm I})$   &   $(160 M_{\rm I})$  &  VAPOR   \\[2pt]

\ \ \ $\left|\vec{v}_{\rm I}\right| = 10$ km/s     &  $2.1 \times 10^{16}$ kg  &  $1.9 \times 10^{17}$ kg   &  $1.2 \times 10^{17}$ kg  &  $7.0 \times 10^{16}$ kg  &  $4.5 \times 10^{16}$ kg  &  $3.1 \times 10^{16}$ kg  &  $2.4 \times 10^{16}$ kg    \\[2pt]
\ \ \ &  $(1.4 \times 10^4 M_{\rm I})$   &   $(1.3 \times 10^5 M_{\rm I})$  &    $(7.9 \times 10^4 M_{\rm I})$                       
&   $(4.7 \times 10^4 M_{\rm I})$    &   $(3.0 \times 10^4 M_{\rm I})$   &   $(2.1 \times 10^4 M_{\rm I})$  &  $(1.6 \times 10^4 M_{\rm I})$   \\[2pt]

\ \ \ $\left|\vec{v}_{\rm I}\right| = 30$ km/s         &  $1.1 \times 10^{16}$ kg   &   $1.0 \times 10^{17}$ kg  &  $5.4 \times 10^{16}$ kg &  $2.8 \times 10^{16}$ kg  &  $1.6 \times 10^{16}$ kg &  $9.7 \times 10^{15}$ kg  &  $6.4 \times 10^{15}$ kg     \\[2pt]
\ \ \ &  $(7300 M_{\rm I})$   &   $(6.7 \times 10^4 M_{\rm I})$  &    $(3.6 \times 10^4 M_{\rm I})$                       
&   $(1.9 \times 10^4 M_{\rm I})$    &   $(1.1 \times 10^4 M_{\rm I})$   &   $(6500 M_{\rm I})$  &  $(4300 M_{\rm I})$   \\[2pt]

\ \ \ $\left|\vec{v}_{\rm I}\right| = 60$ km/s      &  $8.8 \times 10^{15}$ kg  &   $8.4 \times 10^{16}$ kg  &  $4.5 \times 10^{16}$ kg  &  $2.4 \times 10^{16}$ kg  &  $1.4 \times 10^{16}$ kg &  $8.7 \times 10^{15}$ kg &  $5.7 \times 10^{15}$ kg     \\[2pt]
\ \ \ &  $(5900 M_{\rm I})$   &   $(5.6 \times 10^4 M_{\rm I})$  &    $(3.0 \times 10^4 M_{\rm I})$                       
&   $(1.6 \times 10^4 M_{\rm I})$    &   $(9300 M_{\rm I})$   &   $(5800 M_{\rm I})$  &  $(3800 M_{\rm I})$   \\[2pt]

\ \ \ $\left|\vec{v}_{\rm I}\right| = 100$ km/s   &  $7.6 \times 10^{15}$ kg    &  $7.5 \times 10^{16}$ kg   &  $4.0 \times 10^{16}$ kg   &  $2.2 \times 10^{16}$ kg  &  $1.3 \times 10^{16}$ kg   &  $8.1 \times 10^{15}$ kg &  $5.4 \times 10^{15}$ kg  \\[2pt]
\ \ \ &  $(5100 M_{\rm I})$   &   $(5.0 \times 10^4 M_{\rm I})$  &    $(2.7 \times 10^4 M_{\rm I})$                       
&   $(1.5 \times 10^4 M_{\rm I})$    &   $(8600 M_{\rm I})$   &   $(5400 M_{\rm I})$  &  $(3600 M_{\rm I})$   \\[2pt]

\hline
\end{tabular}
\end{minipage}
\end{table*}

A final consideration is the potential destruction of the target due to the impact.
The full-chain analytical model applies only when initial conditions are chosen so that the impactor
does not destroy the
target during the collision. Section 5 of \cite{movetal2016} provided a concise
analytical formulation for when the target would break apart
due to an impactor:

\begin{equation}
\left(\frac{1}{2}\right)\left(\frac{\epsilon M_{\rm T} + M_{\rm I}}{M_{\rm T} + M_{\rm I}}\right)
                        \left(\frac{M_{\rm T} M_{\rm I}}{M_{\rm T} + M_{\rm I}}  \right) \left|\vec{v}_{\rm I}\right|^2
> E_{\rm max}
\label{colcon}
\end{equation}

\noindent{where}

\begin{equation}
\epsilon \equiv \left\{
\begin{array}{ll}
  
  \frac{3 R_{\rm I} l^2 - l^3}{4 R_{\rm I}^3},
  & \quad l < 2 R_{\rm I} \\
  
  1,
  & \quad l \ge 2 R_{\rm I}
\end{array}
\right.
,
\label{auxep}
\end{equation}

\begin{equation}
l \equiv \left(R_{\rm T} + R_{\rm I} \right) \left(1 - \cos{\theta} \right)
,
\label{auxl}
\end{equation}

\noindent{}and

\begin{equation}
E_{\rm max} = \left(13.4 \pm 10.8\right)
\left[\frac{3GM_{\rm T}^2}{5R_{\rm T}} + \frac{3GM_{\rm I}^2}{5R_{\rm I}} + \frac{GM_{\rm T}M_{\rm I}}{R_{\rm T} + R_{\rm I}} \right]
,
\label{Emax}
\end{equation}

\noindent{}where the numerical range given is conservative and encompasses
$\theta = 45^{\circ} - 90^{\circ}$. The relations in equations (\ref{colcon}-\ref{Emax}) 
can be reduced to just two degrees of freedom if a head-on collision of 
equal-composition impactor and target are assumed (see Eq. 21 of \citealt*{veretal2018b}).
For a basaltic impactor and target, here we approximate the destruction condition in that 
equation as

\begin{equation}
\frac{\left|\vec{v}_{\rm I}\right|}{v_{\rm esc}}
\gtrsim
\sqrt{3\left(\frac{R_{\rm T}}{R_{\rm I}}\right)^3}
\label{destroy}
\end{equation}

\noindent{}and use equation (\ref{destroy}) to determine when a target would be destroyed.


\subsection{Simulation data}

\begin{figure}
\includegraphics[width=8cm]{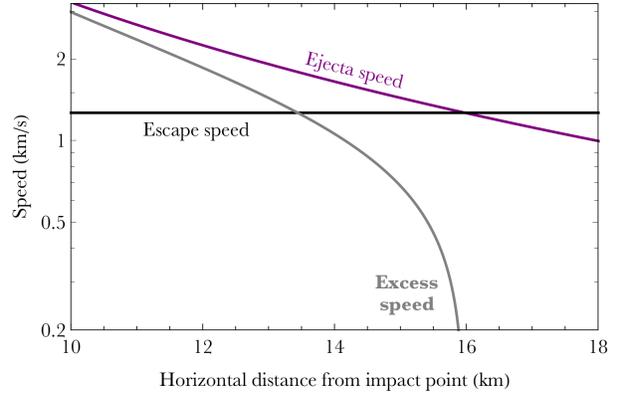}
\caption{
Detail of how the post-collision ejecta speed tapers off as a function of distance
from the impact point, eventually falling below the escape speed of the target body.
The excess speed is the ejecta's speed upon leaving the gravitational pull of the target.
These quantities were computed for $R_{\rm T} = 10^3$ km, $R_{\rm I} = 10^{-4}R_{\rm T}$,
$\left| \vec{v}_{\rm I} \right| = 10$ km/s, $Z=3$ and basaltic
impactors and targets.
}
\label{VelocitiesPlot}
\end{figure}

\begin{figure}
\includegraphics[width=8cm]{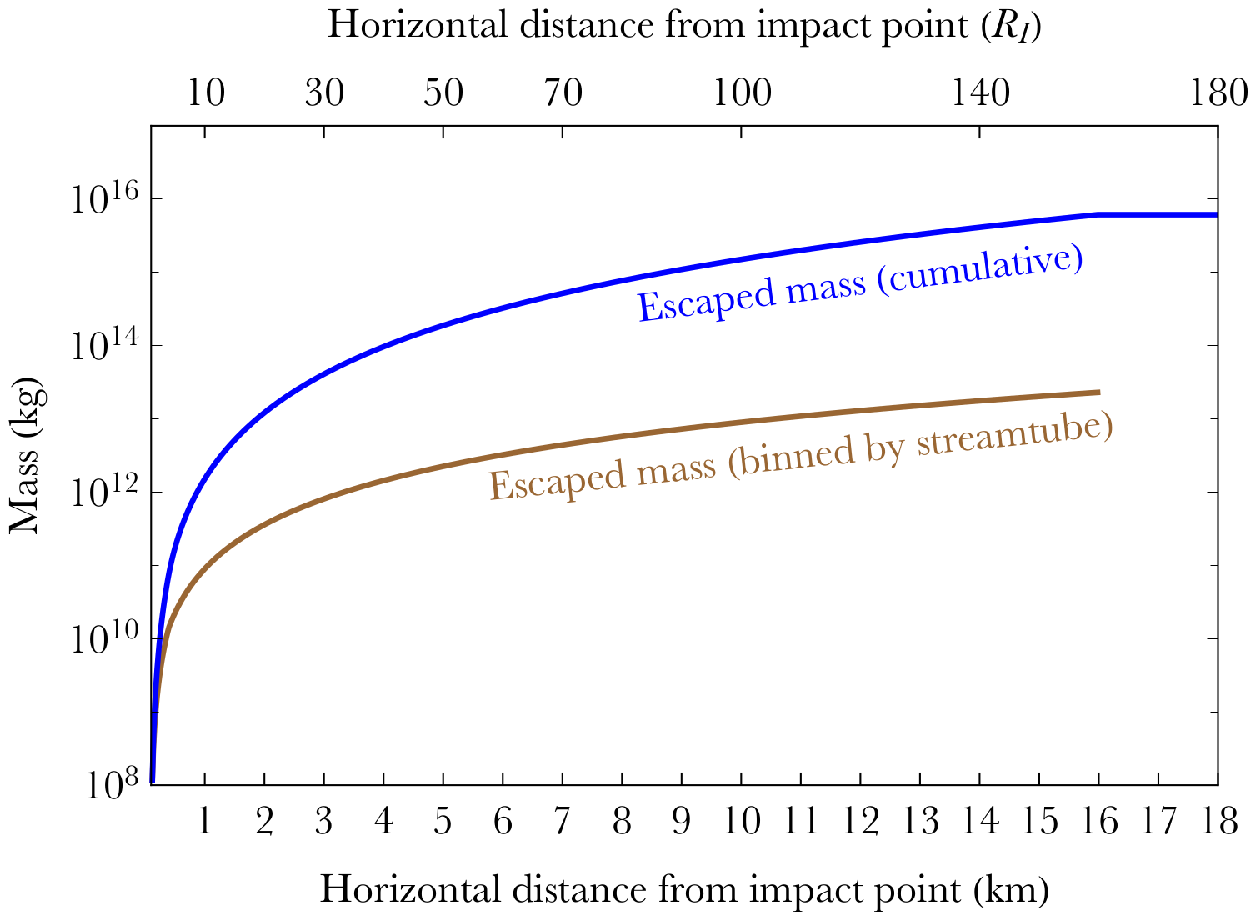}
\caption{
Demonstration of how the amount of mass per bin increases as a function of
horizontal distance from the impact point, for the same impact event that is shown in Fig. \ref{VelocitiesPlot}.
Due to this increase, any mass near the impact point susceptible to vaporization
would be negligible compared to the cumulative mass which escapes the target. 
}
\label{MassTotalPlot}
\end{figure}

\begin{figure}
\includegraphics[width=8cm]{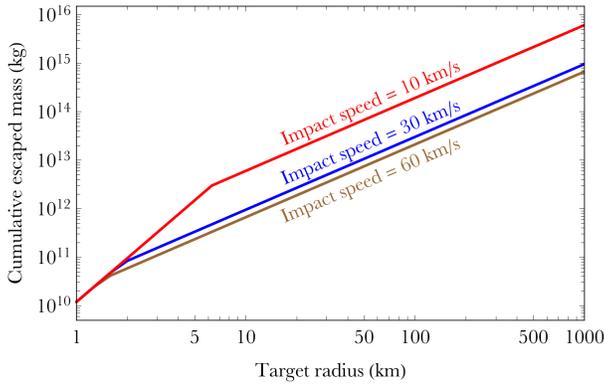}
\caption{
How the magnitude of the escaped mass varies with the target radius and impact speed,
for $R_{\rm I} = 10^{-4}R_{\rm T}$, $Z=3$ and basaltic
impactors and targets.
}
\label{RTPlot}
\end{figure}

\begin{figure}
\includegraphics[width=8cm]{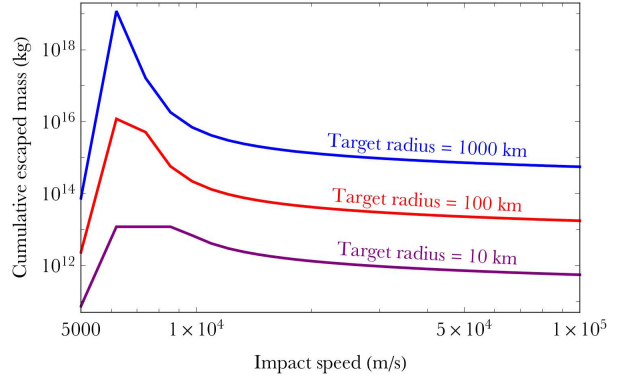}
\caption{
How the magnitude of the escaped mass varies with the impact speed and target radius,
for the same system as in Fig. \ref{RTPlot}. The peak arises partly because of the impact
speed-dependent functional form of $n$ in equation (\ref{npie}).
}
\label{vIPlot}
\end{figure}

\begin{figure}
\includegraphics[width=8cm]{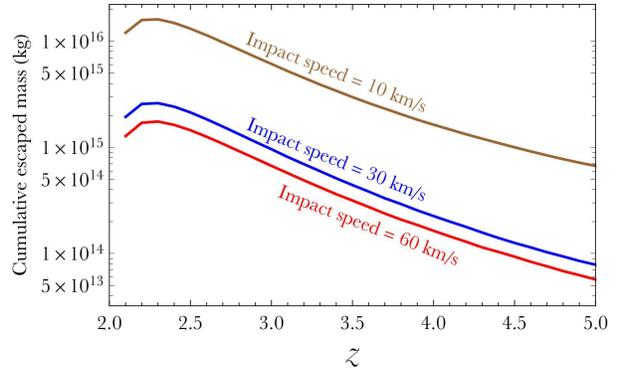}
\caption{
How the magnitude of the escaped mass varies with the entire range of
$Z$ (streamtube shape) and impact speed,
for $R_{\rm T} = 10^3$ km, $R_{\rm I} = 10^{-4}R_{\rm T}$ and basaltic
impactors and targets.
}
\label{ZPlot}
\end{figure}

We present our main results in the form of four data tables 
(Tables \ref{tabfid10m5}-\ref{combos}) and some supplementary
plots of specific cases (Figs. \ref{VelocitiesPlot}-\ref{ZPlot}). Tables  
\ref{tabfid10m5}-\ref{tabfid10m3}
display results for a basaltic impactor and target, where, respectively, 
$R_{\rm I} = 10^{-5}, 10^{-4}, 10^{-3} R_{\rm T}$.

In all tables, the escaped mass is given in terms of $M_{\rm I}$ to two
significant figures.
This presentation (i) enables one to quickly scan the phase space to see where
there is a guaranteed increase of mass in interplanetary space due to a collision
(when the value is greater than unity, which is nearly everywhere),
(ii) keeps the magnitude of the numbers tractable, and (iii) and reflects
the approximate nature of these results, which is appropriate given
the current constraints in white dwarf planetary systems.  


Non-numeric table entries reveal other aspects of these collisions. A value
of ``APART'' indicates likely destruction of the target based on equation (\ref{destroy}).
Values of ``LIMIT'' indicate that the full-chain analytical model cannot approximate
the escaped mass well enough to be used here, based on mass escaping at our limiting location
of $D = 0.2R_{\rm T}$. ``JET'' indicates that equation (\ref{vIcrit}) is not satisfied (such that jetting
and spallation might dominate mass escape),
and ``VAPOR'' indicates that the value of $D$ within which mass escape ceases is under $15R_{\rm I}$ 
(indicating a potential, but still likely low, contribution from vaporized mass).

The presentation of Tables \ref{tabfid10m5}-\ref{tabfid10m3} reflects the strongest dependencies of 
the minimum escaped mass, which are on $\left| \vec{v}_{\rm I} \right|$ and $R_{\rm T}$, and allows one to see how
the result varies over the entire gamut of streamtube shape changes ($Z=2-5$).   
In some cases, particularly in Table \ref{tabfid10m3}, the full-chain
analytical model is applicable for only a subset of $Z$. The diversity of values in Tables \ref{tabfid10m5}-\ref{tabfid10m3}
highlights the danger of oversimplifying the crater ejecta process across the entire phase space with just a single output.

Table \ref{combos} instead displays the magnitude of the minimum escaped mass when the impactor and 
target are made of different materials
for  $R_{\rm T} = 500$~km and $R_{\rm I}= 10^{-3} R_{\rm T} = 5$~km\footnote{A basalt target and impactor is also included here,
because this combination of radii is not sampled elsewhere.}. The table shows that the minimum escaped mass fractions for collisions involving basalt, granite, and iron are roughly within the same order of magnitude. The significant outlier is water, for which the minimum escaped mass appears to be lower, and is the only target material for which the minimum escaped mass does not exceed the impactor mass (but in one case only). 

Figs. \ref{VelocitiesPlot}-\ref{ZPlot} help illustrate the dependencies of various variables on minimum escaped mass for selected impact scenarios. Excavated mass cannot escape the target unless its speed exceeds the escape speed (Fig. \ref{VelocitiesPlot}). The mass which escapes contains an excess speed that is well above the escape speed in all cases except for a narrow band of $D$ (representing about 2.5 km in the figure, or about 15 per cent of the distance to the impact point). The location where the escape and ejecta speeds are equal (at about $D=16$ km) signals the furthest distance from the impact point where material escapes the target. 

For the same system, Fig. \ref{MassTotalPlot} illustrates the mass loss profile. The lower brown line represents the escaped mass binned by streamtube, which monotonically increases until the cutoff point (where material stops escaping). This monotonic increase demonstrates that most of the mass which escapes occurs far from the impact point: at many tens of $R_{\rm I}$ (upper $x$-axis). The consequences are that any vaporization which occurs within a few $R_{\rm I}$ would provide a negligible contribution to the cumulative escaped mass.

Figures \ref{RTPlot} and \ref{vIPlot} reveal largely linear dependencies on log-log scales of the escaped mass on both the target radius and impact speed, at least for the basaltic $R_{\rm I} = 10^{-4} R_{\rm T}$ and $Z=3$ case. However, the dependence of escaped mass on $Z$ is not as obvious, as indicated by Fig. \ref{ZPlot}. The figure illustrates that the escaped mass can vary by several orders of magnitude across all values of $Z$, with the greatest escaping mass occurring for a $Z$ value between 2.0 and 2.5. This trend is largely consistent with the impact scenarios presented across Tables \ref{tabfid10m5}-\ref{combos}.

\section{Observational constraints}

Unless observations of a white dwarf planetary system fortuitously happen to coincide with an intra-system collision \citep{beasok2013,wanetal2019}, what we actually would see is the aftermath. The endstate for much of the debris ejecta -- after being slowly radiatively dragged towards the white dwarf -- will be accretion onto its photosphere. These metal pollution signatures are observed in abundance (see Section 1.1.1). Hence, the salient question is, can the post-collision debris be seen either in place or en route to the white dwarf?

Most dust around white dwarfs is found within the star's disruption, or Roche, radius in the form of compact rings or discs \citep{farihi2016,manetal2020}. These structures do not necessarily arise from radially drifting debris, but instead could be formed by the destruction of exo-asteroids which veer into \citep{graetal1990,jura2003,debetal2012,veretal2014c,malper2020a} or close to \citep{makver2019,veretal2020a} the Roche radius at about 0.005 au. Further, any particles sublimated off a volatile-rich planetesimal would occur around the Roche radius \citep{stoetal2015,veretal2015c,vanetal2018}. Therefore, observationally testing the efficacy of the impact mechanism would require observations of dust which reside at a distance that is beyond several times the white dwarf Roche radius.

Mid-IR observations represent a possible probe of this region because of the wavelengths at which dust emits. The upper limits to 20-30 micron flux observations rule out significant influxes of dust from larger separations, as would be expected for dust being fed by collisions between minor planets. Hence, these observations are important, but unfortunately at this time are sparse. \cite{faretal2014} placed limits on the dust mass in the G29-38 planetary system as a function of dust temperature. They showed that at, for example, a distance of about 11 au, the dust mass cannot be higher than that of Haumea ($3 \times 10^{21}$ kg). Another system with constraints is GD 362: \cite{xuetal2013} established limits on the dust mass ($10^{22} - 10^{26}$ kg) for separations ranging between 5-30 au.

The lack of (a substantial amount of) cold dust in these two systems at distances of several to tens of au suggests that a distant collision event with ejecta mass exceeding that of the largest minor planets in the solar system has not occurred recently (relative to the white dwarf cooling ages). However, if metal pollution primarily arises from bodies smaller than Haumea, then these mass limits might be too high to produce useful constraints. These systems also represent only about 0.2 per cent of all known polluted white dwarf systems. Therefore, although we cannot make any sweeping conclusions about the efficacy of the impact mechanism for producing pollutants by using these constraints, we certainly appeal for additional observations of cold dust in polluted white dwarf planetary systems.

\section{Possible scenarios}

Collisions can occur anywhere within white dwarf planetary systems: both near the white dwarf Roche radius at 0.005 au and at distances of 100 au. The computations we have provided here in Tables \ref{tabfid10m5}-\ref{combos} can be applied across this entire range. Now we outline two possible scenarios in which the results of this paper may be used. 

\subsection{Cool white dwarf with asteroid on compact orbit}

The minor planet or planets orbiting WD 1145+017 are detectable because they are close enough to the white dwarf (with a separation of approximately 0.005 au) to be breaking up and producing transiting debris. If, however, a $10^3$ km-radius minor planet was slightly further away from the white dwarf, at a distance of say, 0.5 au (similar to the semimajor axis of the debris orbiting ZTF J0139+5245; \citealt*{vanetal2019}), then it would not be disrupted and could be hidden from view \citep{vanvan2018}. 

This $10^3$ km-radius minor planet would be too large to be radiatively dragged towards the white dwarf, which we take to have a cooling age of 3 Gyr (similar to the age of LSPM J0207+3331, which has reported infrared excesses; \citealt*{debetal2019}). Hence, this large minor planet is dynamically stagnant: not detectable, and not polluting the white dwarf.

Suppose further that the mass reservoirs in the outer planetary system (say at around 5 au) are not suited to produce the type of dynamical encounters envisaged by the canonical model for pollution:  In other words,  the major planet or planets at this distance cannot perturb a pool of smaller 1 km-radius asteroids (at around 5 au) onto orbits which graze the white dwarf Roche radius (at 0.005 au), nor even reach three times this distance, where breakup could occur \citep{makver2019,veretal2020a}. The eccentricity required to reach 0.015 au from 5 au would be about 0.997.

Failure to achieve this eccentricity would mean that the white dwarf would never be metal polluted. However, if these small 1 km-radius minor planets at 5 au are perturbed more gently, to achieve significantly less eccentric orbits (with eccentricities of just about 0.9), then they could collide with the $10^3$ km-radius minor planet at 0.5 au. 

If we assume that both the target and impactor are basaltic, then we can use Table \ref{tabfid10m3} to characterize the collision. We estimate that the collision speed is about 60 km/s from equation (\ref{maxorb}), meaning that we would read across the corresponding row in the bottommost section of Table \ref{tabfid10m3}.

We find that depending on the geometry adopted, about $10^{17} - 10^{18}$ kg of ejecta mass would be produced in a single collision. This amount well exceeds the impactor mass by many orders of magnitude. This ejecta mass is now available to pollute the white dwarf. Even though the white dwarf is cool at 3 Gyr old, its radiation would drag inwards dust and pebbles at 0.5 au, even those on highly eccentric orbits. Although the pollution timescale would be a detailed function of both the geometry of the collision and the size distribution of the crater ejecta, we can estimate from \cite{veretal2015b} that the debris would pollute the white dwarf within a few orders of magnitude of Myrs. Also, during this period, other collisions with the original target might occur, making available additional potentially polluting debris.

\subsection{Hot white dwarf with asteroid on wide orbit}

Now consider a completely different scenario. Suppose a hot white dwarf has the same 13 Myr cooling age as WD J0914+1914 \citep{ganetal2019} and hosts a series of planets and asteroids which all have semimajor axes and orbital pericentres beyond 10 au. Hence the region between the white dwarf photosphere and 10 au is clear of material. Further assume that there is no future prospect of the minor planets being perturbed within this boundary.

In this case, the white dwarf would never be polluted, and also give no indication of the planetary system which it hosts. Suppose, however, that a 500 km-radius iron core resides at a distance of 20 au and is smashed into by a collection of 0.5 km-radius basaltic impactors. Hence, we use Table \ref{combos} to model these collisions. At this separation, a likely collision speed is 5 km/s. For this speed, the bottommost portion of the table indicates that while some of the ejecta is vaporized, a single collision will generally produce about $10^{14} - 10^{15}$ kg of debris.

This debris is now available to pollute the white dwarf. Because the white dwarf is so young and luminous (with a luminosity of about a tenth of the Sun's), it can drag this debris into its photosphere, even from distances of tens of au. The drag timescale is again dependent on the geometry of the collision and the size distribution assumed for the debris, but roughly can occur on Myr timescales.

\section{Discussion}

The results in this paper can be used to help constrain dynamical models of planetary system evolution which lead to polluted white dwarfs. Despite the mounting observations of planetary debris in a white dwarf's immediate circumstellar environment (see Section 1.1), understanding the dynamical history of these systems remains an important but largely outstanding task. Current treatments which involve only {\rev dynamical interactions between} major and minor planets as pollution sources may be too simplistic.

Debris in the interplanetary environment represents another pathway for metal pollution, particularly in cases when exo-asteroids themselves are rarely perturbed towards the white dwarf, or perturbed towards the white dwarf only in concentrated periods of time (which are unlikely to be observed). As shown here, a net gain of interplanetary debris can easily be generated by a single collision on an atmosphere-less target; these targets are likely to be common in white dwarf planetary systems subsequent to the evaporative effects of the progenitor star. 

Debris-producing collisions can be generated from gravitational instabilities amongst the extant objects (which likely involve at least one major planet) on short or long timescales, as well as from stellar flybys. Hence, considering impact debris in white dwarf planetary systems which experience bombardments from Nice-model like events may be important. Subsequently, an analysis of the impact geometry, escape geometry, and speed of the debris coupled with gravitational perturbations and radiative drag forces can yield time-dependent mass and size distributions, as well as chemical composition information based on impactor and target materials.

This information may be fed into various models, such as, for example, those from \cite{wyaetal2014}, \cite{broetal2017} and \cite{kenbro2017a,kenbro2017b}, to obtain a more accurate population synthesis. \cite{wyaetal2014} constrained the mass distribution of accreted material from observed atmospheric pollution, and considered stochastic and continuous accretion regimes. \cite{broetal2017} provided formulae which determine if debris that is steeply infalling into a white dwarf will be sublimated, fragmented or impacted onto the stellar photosphere. \cite{kenbro2017a} and \cite{kenbro2017b} utilised collisional cascades to model white dwarf debris discs, and incorporated inflow from the circumstellar environment. 

Establishing a compositional link between crater ejecta and metals found in white dwarfs is another important extension of these results. If observational distinctions between detections of the bulk remains of an exo-asteroid versus the crustal material of an exo-planet can be enhanced {\rev \citep{haretal2018,holetal2018,bonetal2020}}, then understanding the mass and dynamical pathways of impact ejecta would become more important. Tracing the chemistry and dynamics back to planetary system formation along the early main sequence phase {\rev \citep{haretal2018,bonetal2020}} has the potential to link planetary architecture with instabilities, bombardments and observed accretion rates.

The impact crater formalism developed in \cite{kurtak2019} and used here provides a fully analytical alternative to the $\pi$-group scaling laws \citep{holsch1982}, which require resource-consuming experimental work to obtain empirical constants across the phase space. The flexibility and speed of the \cite{kurtak2019} formalism facilitates the study of white dwarf planetary systems because of their potential to harbour a wide variety of objects and architectures.  Extending the formalism to account for very high speed impacts ($> 100$ km/s) and objects of high internal strength could then be applicable to impacts which occur just outside the white dwarf Roche radius, and for objects like the high-strength exo-asteroid discovered by \cite{manetal2019}. Because tidal interactions can quickly alter orbits {\rev \citep{veretal2019b,verwol2019,verful2019,verful2020}}, the same target might experience different impact regimes over short timescales relative to the white dwarf cooling age. 
 
Further, if the magnitude of the vaporized mass represented a non-negligible fraction of the escaped ejecta mass, then the temperature of the target and the white dwarf cooling age would have been important considerations. However, the independence of these variables suggest that the results in this paper may be applied to exoplanetary systems with any type of star. The implications of debris generation for giant branch planetary systems are just as important, particularly for the material which survives to the white dwarf phase (and is eventually observable there).

\section{Summary}

We have quantitatively introduced the idea that metal pollution in some white dwarf atmospheres may be due in part or in full to post-impact debris rather than solely the ingestion of entire asteroids, comets or moons. Hence, white dwarf planetary systems where minor planets do not have a dynamical pathway to reach the central star themselves can nevertheless feature a polluted white dwarf because of collisional ejecta following bombardment events. We utilised the full-chain analytical impact crater model of \cite{kurtak2019} to compute lower bounds on the ejecta mass which escapes the gravitational pull of an atmosphere-less target subsequent to a single collision. Our applications spanned a range of impact speeds (5-100 km/s), target radii ($10^{0-3}$ km) and chemical compositions (basalt, granite, water, iron) which are either known or expected to exist in white dwarf planetary systems. We report our minimum escape estimates in Tables \ref{tabfid10m5}-\ref{combos}, which provide values that can constrain future dynamical modelling studies. Our results may be applied within any exo-planetary system, {\rev not just those containing white dwarfs}.

\section*{Acknowledgements}

We thank an anonymous reviewer and Boris Ivanov for valuable comments which have greatly improved the manuscript. DV gratefully acknowledges the support of the STFC via an Ernest Rutherford Fellowship (grant ST/P003850/1), and funding from the University of Warwick's Habitability Global Research Priorities (GRP) Programme. KK is supported by JSPS KAKENHI grant numbers JP17H01176, JP17H01175, JP17K18812, JP17H02990, JP18HH04464, JP19H00726 and by the Astrobiology Center of the National Institute of Natural Sciences, NINS (AB301018). We also acknowledge useful initial discussions at the 9th Workshop on Catastrophic Disruption in the Solar System in Kobe.

\label{lastpage}

\begin{thebibliography}{99}


\bibitem[Ahrens \& O'Keefe(1972)]{ahroke1972} Ahrens, T.~J., \& O'Keefe, J.~D.\ 1972, The Moon, 4, 214

\bibitem[Alcock et al.(1986)]{alcetal1986} Alcock, C., Fristrom, C.~C., \& Siegelman, R.\ 1986, ApJ, 302, 462 

\bibitem[Anderson et al.(2003)]{andetal2003} Anderson, J.~L.~B., Schultz, P.~H., \& Heineck, J.~T.\ 2003, Journal of Geophysical Research (Planets), 108, 5094

\bibitem[Anderson et al.(2004)]{andetal2004} Anderson, J.~L.~B., Schultz, P.~H., \& Heineck, J.~T.\ 2004, Meteoritics and Planetary Science, 39, 30

\bibitem[Andrews(2020)]{andrews2020} Andrews, S.~M.\ 2020, ARA\&A In Press, arXiv:2001.05007

\bibitem[Ang(1990)]{ang1990} Ang, J.~A.\ 1990, International Journal of Impact Engineering, 10, 23.

\bibitem[Artemieva \& Ivanov(2004)]{artiva2004} Artemieva, N., \& Ivanov, B.\ 2004, Icarus, 171, 84 

\bibitem[Artemieva \& Shuvalov(2008)]{artshu2008} Artemieva, N.~A., \& Shuvalov, V.~V.\ 2008, Solar System Research, 42, 329.

\bibitem[Bear \& Soker(2011)]{beasok2011} Bear, E., \& Soker, N.\ 2011, MNRAS, 414, 1788 

\bibitem[Bear \& Soker(2013)]{beasok2013} Bear, E., \& Soker, N.\ 2013, New Astronomy, 19, 56 

\bibitem[Bear \& Soker(2015)]{beasok2015} Bear, E., \& Soker, N.\ 2015, MNRAS, 450, 4233

\bibitem[Bierhaus et al.(2013)]{bieetal2013} Bierhaus M., Noack L., W{\"u}nnemann K., Breuer D., 2013, LPSC, 2420

\bibitem[Bodman et al.(2018)]{bodetal2018} Bodman, E.~H.~L., Wright, J.~T., Desch, S.~J., et al.\ 2018, AJ, 156, 173

\bibitem[Bonsor \& Wyatt(2010)]{bonwya2010} Bonsor, A., \& Wyatt, M.\ 2010, MNRAS, 409, 1631

\bibitem[Bonsor et al.(2013)]{bonetal2013} Bonsor, A., Kennedy, G.~M., Crepp, J.~R., et al.\ 2013, MNRAS, 431, 3025 

\bibitem[Bonsor et al.(2014)]{bonetal2014} Bonsor, A., Kennedy, G.~M., Wyatt, M.~C., Johnson, J.~A., \& Sibthorpe, B.\ 2014, MNRAS, 437, 3288 

\bibitem[Bonsor \& Veras(2015)]{bonver2015} Bonsor, A., \& Veras, D.\ 2015, MNRAS, 454, 53 

\bibitem[Bonsor et al.(2020)]{bonetal2020} Bonsor, A., Carter, P.~J., Hollands, M., et al.\ 2020, MNRAS, 492, 2683

\bibitem[Brown et al.(2017)]{broetal2017} Brown, J.~C., Veras, D., \& G{\"a}nsicke, B.~T.\ 2017, MNRAS, 468, 1575 

\bibitem[Burchell \& Mackay(1998)]{burmac1998} Burchell, M.~J., \& Mackay, N.~G.\ 1998, JGR, 103, 22761

\bibitem[Cabral et al.(2019)]{cabetal2019} Cabral, N., Lagarde, N., Reyl{\'e}, C., et al.\ 2019, A\&A, 622, A49

\bibitem[Caiazzo \& Heyl(2017)]{caihey2017} Caiazzo, I., \& Heyl, J.~S.\ 2017, MNRAS, 469, 2750

\bibitem[Campbell et al.(1988)]{cametal1988} Campbell, B., Walker, G.~A.~H., \& Yang, S.\ 1988, ApJ, 331, 902 

\bibitem[Cauley et al.(2018)]{cauetal2018} Cauley, P.~W., Farihi, J., Redfield, S., et al.\ 2018, ApJL, 852, L22

\bibitem[Cintala et al.(1999)]{cinetal1999} Cintala, M.~J., Berthoud, L., \& H{\"o}rz, F.\ 1999, Meteoritics and Planetary Science, 34, 605.

\bibitem[Col{\'o}n et al.(2018)]{coletal2018} Col{\'o}n, K.~D., Zhou, G., Shporer, A., et al.\ 2018, AJ, 156, 227

\bibitem[Coutu et al.(2019)]{couetal2019} Coutu, S., Dufour, P., Bergeron, P., et al.\ 2019, arXiv:1907.05932

\bibitem[Croft(1980)]{croft1980} Croft, S.~K.\ 1980, Lunar and Planetary Science Conference Proceedings, 3, 2347.

\bibitem[de Sousa Ribeiro et al.(2020)]{desetal2020} de Sousa Ribeiro, R., Morbidelli, A., Raymond, S.~N., Izidoro, A., Gomes R., Vieira Neto, E.\ 2020, Icarus, 339, 113605

\bibitem[Debes \& Sigurdsson(2002)]{debsig2002} Debes, J.~H., \& Sigurdsson, S.\ 2002, ApJ, 572, 556 

\bibitem[Debes et al.(2012)]{debetal2012} Debes, J.~H., Walsh, K.~J., \& Stark, C.\ 2012, ApJ, 747, 148

\bibitem[Debes et al.(2019)]{debetal2019} Debes, J.~H., Th{\'e}venot, M., Kuchner, M.~J., et al.\ 2019, ApJL, 872, L25

\bibitem[DeCarli et al.(2007)]{decetal2007} DeCarli, P.~S., El Goresy, A., Xie, Z., et al.\ 2007, Shock Compression of Condensed Matter, 1371.

\bibitem[DeCarli(2013)]{decarli2013} DeCarli, P.~S.\ 2013, Proc. Eng. 58, 570

\bibitem[Dong et al.(2010)]{donetal2010} Dong, R., Wang, Y., Lin, D.~N.~C., \& Liu, X.-W.\ 2010, ApJ, 715, 1036 

\bibitem[Doyle et al.(2019)]{doyetal2019} Doyle, A.~E., Young, E.~D., Klein, B., et al.\ 2019, Science, 366, 356

\bibitem[Dufour et al.(2007)]{dufetal2007} Dufour, P., Bergeron, P., Liebert, J., et al.\ 2007, ApJ, 663, 1291

\bibitem[Dufour et al.(2012)]{dufetal2012} Dufour, P., Kilic, M., Fontaine, G., et al.\ 2012, ApJ, 749, 6

\bibitem[Farihi et al.(2013)]{faretal2013} Farihi, J., G{\"a}nsicke, B.~T., \& Koester, D.\ 2013, Science, 342, 218 

\bibitem[Farihi et al.(2014)]{faretal2014} Farihi, J., Wyatt, M.~C., Greaves, J.~S., et al.\ 2014, MNRAS, 444, 1821

\bibitem[Farihi(2016)]{farihi2016} Farihi, J.\ 2016, New Astronomy Reviews, 71, 9 

\bibitem[Fortin-Archambault et al.(2020)]{foretal2020} Fortin-Archambault, M., Dufour, P., \& Xu, S.\ 2020, ApJ, 888, 47

\bibitem[G{\"a}nsicke et al.(2012)]{gaeetal2012} G{\"a}nsicke, B.~T., Koester, D., Farihi, J., et al.\ 2012, MNRAS, 424, 333 

\bibitem[G{\"a}nsicke et al.(2019)]{ganetal2019} G{\"a}nsicke, B.~T., Schreiber, M.~R., Toloza, O., et al.\ 2019, Nature, 576, 61

\bibitem[Gentile Fusillo et al.(2015)]{genetal2015} Gentile Fusillo, N.~P., G{\"a}nsicke, B.~T., \& Greiss, S.\ 2015, MNRAS, 448, 2260

\bibitem[Gentile Fusillo et al.(2017)]{genetal2017} Gentile Fusillo, N.~P., G{\"a}nsicke, B.~T., Farihi, J., et al.\ 2017, MNRAS, 468, 971 

\bibitem[Goldstein(1987)]{goldstein1987} Goldstein, J.\ 1987, A\&A, 178, 283 

\bibitem[Gomes et al.(2005)]{gometal2005} Gomes, R., Levison, H.~F., Tsiganis, K., \& Morbidelli, A.\ 2005, Nature, 435, 466 

\bibitem[Graham et al.(1990)]{graetal1990} Graham, J.~R., Matthews, K., Neugebauer, G., et al.\ 1990, ApJ, 357, 216

\bibitem[Greenstreet et al.(2015)]{greetal2015} Greenstreet, S., Gladman, B., \& McKinnon, W.~B.\ 2015, Icarus, 258, 267.

\bibitem[Grey et al.(2002)]{greetal2002} Grey, I.~D.~S., Burchell, M.~J., \& Shrine, N.~R.~G.\ 2002, Journal of Geophysical Research (Planets), 107, 5076

\bibitem[Grishin \& Veras(2019)]{griver2019} Grishin, E., \& Veras, D.\ 2019, MNRAS, 489, 168

\bibitem[Hamers \& Portegies Zwart(2016)]{hampor2016} Hamers, A.~S., \& Portegies Zwart, S.~F.\ 2016, MNRAS, 462, L84 

\bibitem[Harrison et al.(2018)]{haretal2018} Harrison, J.~H.~D., Bonsor, A., \& Madhusudhan, N.\ 2018, MNRAS, 479, 3814 

\bibitem[Head et al.(2002)]{heaetal2002} Head, J.~N., Melosh, H.~J., \& Ivanov, B.~A.\ 2002, Science, 298, 1752.

\bibitem[Hinkel \& Unterborn(2018)]{hinunt2018} Hinkel, N.~R., \& Unterborn, C.~T.\ 2018, ApJ, 853, 83

\bibitem[Hollands et al.(2017)]{holetal2017} Hollands, M.~A., Koester, D., Alekseev, V., Herbert, E.~L., \& G{\"a}nsicke, B.~T.\ 2017, MNRAS, 467, 4970 

\bibitem[Hollands et al.(2018)]{holetal2018} Hollands, M.~A., G{\"a}nsicke, B.~T., \& Koester, D.\ 2018, MNRAS, 477, 93 

\bibitem[Holsapple \& Schmidt(1982)]{holsch1982} Holsapple, K.~A., \& Schmidt, R.~M.\ 1982, JGR, 87, 1849 

\bibitem[Housen et al.(1983)]{houetal1983} Housen, K.~R., Schmidt, R.~M., \& Holsapple, K.~A.\ 1983, Journal of Geophysical Research, 88, 2485.

\bibitem[Housen \& Holsapple(2011)]{houhol2011} Housen, K.~R., \& Holsapple, K.~A.\ 2011, Icarus, 211, 856.

\bibitem[Ito \& Malhotra(2006)]{itomal2006} Ito, T., \& Malhotra, R.\ 2006, Advances in Space Research, 38, 817.

\bibitem[Johnson et al.(2014)]{johetal2014} Johnson, B.~C., Bowling, T.~J., \& Melosh, H.~J.\ 2014, Icarus, 238, 13.

\bibitem[Johnson et al.(2015)]{johetal2015} Johnson, B.~C., Minton, D.~A., Melosh, H.~J., et al.\ 2015, Nature, 517, 339.

\bibitem[Jura(2003)]{jura2003} Jura, M.\ 2003, ApJL, 584, L91

\bibitem[Jura et al.(2012)]{juretal2012} Jura, M., Xu, S., Klein, B., Koester, D., \& Zuckerman, B.\ 2012, ApJ, 750, 69 

\bibitem[Jura \& Young(2014)]{juryou2014} Jura, M., \& Young, E.~D.\ 2014, Annual Review of Earth and Planetary Sciences, 42, 45 

\bibitem[Kepler et al.(2015)]{kepetal2015} Kepler, S.~O., Pelisoli, I., Koester, D., et al.\ 2015, MNRAS, 446, 4078 

\bibitem[Kepler et al.(2016)]{kepetal2016} Kepler, S.~O., Pelisoli, I., Koester, D., et al.\ 2016, MNRAS, 455, 3413

\bibitem[Kenyon \& Bromley(2017a)]{kenbro2017a} Kenyon, S.~J., \& Bromley, B.~C.\ 2017a, ApJ, 844, 116 

\bibitem[Kenyon \& Bromley(2017b)]{kenbro2017b} Kenyon, S.~J., \& Bromley, B.~C.\ 2017b, ApJ, 850, 50 

\bibitem[Kieffer(1977)]{kieffer1977} Kieffer, S.~W.\ 1977, Impact and Explosion Cratering: Planetary and Terrestrial Implications, 751.

\bibitem[Klein et al.(2010)]{kleetal2010} Klein, B., Jura, M., Koester, D., Zuckerman, B., \& Melis, C.\ 2010, ApJ, 709, 950 

\bibitem[Klein et al.(2011)]{kleetal2011} Klein, B., Jura, M., Koester, D., \& Zuckerman, B.\ 2011, ApJ, 741, 64 

\bibitem[Kleinman et al.(2013)]{kleetal2013} Kleinman, S.~J., Kepler, S.~O., Koester, D., et al.\ 2013, ApJS, 204, 5

\bibitem[Koester(2009)]{koester2009} Koester, D.\ 2009, A\&A, 498, 517 
  
\bibitem[Koester et al.(2014)]{koeetal2014} Koester, D., G{\"a}nsicke, B.~T., \& Farihi, J.\ 2014, A\&A, 566, A34 

\bibitem[Kostov et al.(2016)]{kosetal2016} Kostov, V.~B., Moore, K., Tamayo, D., Jayawardhana, R., \& Rinehart, S.~A.\ 2016, ApJ, 832, 183 

\bibitem[Kozakis et al.(2018)]{kozetal2018} Kozakis, T., Kaltenegger, L., \& Hoard, D.~W.\ 2018, ApJ, 862, 69

\bibitem[Kral et al.(2018)]{kraetal2018} Kral, Q., Clarke, C., \& Wyatt, M.~C.\ 2018, Handbook of Exoplanets, 165

\bibitem[Kratter \& Perets(2012)]{kraper2012} Kratter, K.~M., \& Perets, H.~B.\ 2012, ApJ, 753, 91 

\bibitem[Krijt et al.(2017)]{krietal2017} Krijt, S., Bowling, T.~J., Lyons, R.~J., et al.\ 2017, ApJ, 839, L21.

\bibitem[Kunitomo et al.(2018)]{kunetal2018} Kunitomo, M., Guillot, T., Ida, S., et al.\ 2018, A\&A, 618, A132

\bibitem[Kurosawa et al.(2015)]{kuretal2015} Kurosawa, K., Nagaoka, Y., Senshu, H., et al.\ 2015, Journal of Geophysical Research (Planets), 120, 1237.

\bibitem[Kurosawa et al.(2018)]{kuretal2018} Kurosawa, K., Okamoto, T., \& Genda, H.\ 2018, Icarus, 301, 219.

\bibitem[Kurosawa \& Takada(2019)]{kurtak2019} Kurosawa, K. \& Takada, S.\ 2019, Icarus, 317, 135

\bibitem[Lingam \& Loeb(2017)]{linloe2017} Lingam, M., \& Loeb, A.\ 2017, Proceedings of the National Academy of Science, 114, 6689.

\bibitem[Liu et al.(2018)]{liuetal2018} Liu, F., Yong, D., Asplund, M., et al.\ 2018, A\&A, 614, A138

\bibitem[Livio \& Soker(1984)]{livsok1984} Livio, M., \& Soker, N.\ 1984, MNRAS, 208, 763 

\bibitem[Makarov \& Veras(2019)]{makver2019} Makarov, V. \& Veras, D.\ 2019, ApJ, 886, 127

\bibitem[Malamud \& Perets(2016)]{malper2016} Malamud, U., \& Perets, H.~B.\ 2016, ApJ, 832, 160 

\bibitem[Malamud \& Perets(2017a)]{malper2017a} Malamud, U., \& Perets, H.~B.\ 2017a, ApJ, 842, 67 

\bibitem[Malamud \& Perets(2017b)]{malper2017b} Malamud, U., \& Perets, H.~B.\ 2017b, ApJ, 849, 8 

\bibitem[Malamud \& Perets(2020a)]{malper2020a} Malamud, U., \& Perets, H.~B.\ 2020a, MNRAS 492, 5561

\bibitem[Malamud \& Perets(2020b)]{malper2020b} Malamud, U., \& Perets, H.~B.\ 2020b, MNRAS In Press, arXiv:1911.12184

\bibitem[Manser et al.(2019)]{manetal2019} Manser, C.~J. et al.\ 2019, Science, 364, 66

\bibitem[Manser et al.(2020)]{manetal2020} Manser, C.~J., G{\"a}nsicke, B.~T., Gentile Fusillo, N.~P., et al.\ 2020, MNRAS In Press, arXiv:2002.01936

\bibitem[Marchi et al.(2014)]{maretal2014} Marchi S., Bottke W.~F., Elkins-Tanton L.~T., Bierhaus M., Wuennemann K., Morbidelli A., Kring D.~A., 2014, Natur, 511, 578

\bibitem[Marcus(1969)]{marcus1969} Marcus, A.~H.\ 1969, Icarus, 11, 76.

\bibitem[Martin et al.(2020)]{maretal2020} Martin, R.~G., Livio, M., Smallwood, J.~L., et al.\ 2020, MNRAS In Press, arXiv:2002.04751

\bibitem[Maxwell(1977)]{maxwell1977} Maxwell, D.~E.\ 1977, In: Impact and Explosion Cratering. Eds: Roddy, D.J., Pepin, R.O., Merrill, R.B., Pergamon Press, Printed in the USA, pgs. 1003-1008

\bibitem[Melis \& Dufour(2017)]{melduf2017} Melis, C., \& Dufour, P.\ 2017, ApJ, 834, 1 

\bibitem[Melosh(1984)]{melosh1984} Melosh, H.~J.\ 1984, Icarus, 59, 234.

\bibitem[Melosh(1985a)]{melosh1985a} Melosh, H.~J.\ 1985a, Geology, 13, 144.

\bibitem[Melosh(1985b)]{melosh1985b} Melosh, H.~J.\ 1985b, Icarus, 62, 339.

\bibitem[Melosh \& Sonett(1986)]{melson1986} Melosh, H.~J., \& Sonett, C.~P.\ 1986, Origin of the Moon, 621.

\bibitem[Melosh(1988)]{melosh1988} Melosh, H.~J.\ 1988, Nature, 332, 687.

\bibitem[Melosh(1989)]{melosh1989} Melosh, H.~J.\ 1989, New York : Oxford University Press ; Oxford : Clarendon Press

\bibitem[Melosh(2003)]{melosh2003} Melosh, H.~J.\ 2003, Astrobiology, 3, 207.

\bibitem[Melosh(2011)]{melosh2011} Melosh, H.~J.\ 2011 Planetary Surface Processes, Cambridge University Press, Cambridge UK. 

\bibitem[Morbidelli et al.(2018)]{moretal2018} Morbidelli, A., Nesvorny, D., Laurenz, V., et al.\ 2018, Icarus, 305, 262 

\bibitem[Movshovitz et al.(2016)]{movetal2016} Movshovitz, N., Nimmo, F., Korycansky, D.~G., Asphaug, E., \& Owen, J.~M.\ 2016, Icarus, 275, 85 

\bibitem[Mustill et al.(2013)]{musetal2013} Mustill, A.~J., Marshall, J.~P., Villaver, E., et al.\ 2013, MNRAS, 436, 2515

\bibitem[Mustill et al.(2014)]{musetal2014} Mustill, A.~J., Veras, D., \& Villaver, E.\ 2014, MNRAS, 437, 1404 

\bibitem[Mustill et al.(2018)]{musetal2018} Mustill, A.~J., Villaver, E., Veras, D., G{\"a}nsicke, B.~T., \& Bonsor, A.\ 2018, MNRAS, 476, 3939 

\bibitem[Nelemans \& Tauris(1998)]{neltau1998} Nelemans, G., \& Tauris, T.~M.\ 1998, A\&A, 335, L85 

\bibitem[O'Brien \& Sykes(2011)]{obrsyk2011} O'Brien, D.~P., \& Sykes, M.~V.\ 2011, Space Science Reviews, 163, 41.

\bibitem[O'Keefe \& Ahrens(1977)]{okeahr1977} Okeefe, J.~D., \& Ahrens, T.~J.\ 1977, Science, 198, 1249.

\bibitem[Paquette et al.(1986)]{paqetal1986} Paquette, C., Pelletier, C., Fontaine, G., \& Michaud, G.\ 1986, ApJS, 61, 197 

\bibitem[Parriott \& Alcock(1998)]{paralc1998} Parriott, J., \& Alcock, C.\ 1998, ApJ, 501, 357 

\bibitem[Payne et al.(2016)]{payetal2016} Payne, M.~J., Veras, D., Holman, M.~J., G\"{a}nsicke, B.~T.\ 2016, MNRAS, 457, 217 

\bibitem[Payne et al.(2017)]{payetal2017} Payne, M.~J., Veras, D., G{\"a}nsicke, B.~T., \& Holman, M.~J.\ 2017, MNRAS, 464, 2557 

\bibitem[Petrovich \& Mu{\~n}oz(2017)]{petmun2017} Petrovich, C., \& Mu{\~n}oz, D.~J.\ 2017, ApJ, 834, 116 

\bibitem[Pierazzo et al.(1997)]{pieetal1997} Pierazzo, E., Vickery, A.~M., \& Melosh, H.~J.\ 1997, Icarus, 127, 408

\bibitem[Pierazzo \& Melosh(2000a)]{piemel2000a} Pierazzo, E., \& Melosh, H.~J.\ 2000a, Icarus, 145, 252

\bibitem[Pierazzo \& Melosh(2000b)]{piemel2000b} Pierazzo, E., \& Melosh, H.~J.\ 2000b, Meteoritics and Planetary Science, 35, 117

\bibitem[Polanskey \& Ahrens(1990)]{polahr1990} Polanskey, C.~A., \& Ahrens, T.~J.\ 1990, Icarus, 87, 140.

\bibitem[Portegies Zwart(2013)]{portegieszwart2013} Portegies Zwart, S.\ 2013, MNRAS, 429, L45 

\bibitem[Raddi et al.(2015)]{radetal2015} Raddi, R., G{\"a}nsicke, B.~T., Koester, D., et al.\ 2015, MNRAS, 450, 2083 

\bibitem[Reach et al.(2005)]{reaetal2005} Reach, W.~T., Kuchner, M.~J., von Hippel, T., et al.\ 2005, ApJL, 635, L161

\bibitem[Reach et al.(2009)]{reaetal2009} Reach, W.~T., Lisse, C., von Hippel, T., et al.\ 2009, ApJ, 693, 697

\bibitem[Ridden-Harper et al.(2019)]{ridetal2019} Ridden-Harper, A.~R., Snellen, I.~A.~G., Keller, C.~U., et al.\ 2019, Submitted to A\&A, arXiv:1906.08795

\bibitem[Santos et al.(2017)]{sanetal2017} Santos, N.~C., Adibekyan, V., Dorn, C., et al.\ 2017, A\&A, 608, A94

\bibitem[Schatzman(1958)]{schatzman1958} Schatzman, E.~L.\ 1958, Amsterdam, North-Holland Pub.~Co.; New York, Interscience Publishers, 1958  

\bibitem[Schreiber et al.(2019)]{schetal2019} Schreiber, M.~R., G{\"a}nsicke, B.~T., Toloza, O., et al.\ 2019, ApJL, 887, L4

\bibitem[Schr{\"o}der \& Smith(2008)]{schcon2008} Schr{\"o}der, K.-P., \& Smith, R.\ 2008, MNRAS, 386, 155 

\bibitem[Schultz(1996)]{schultz1996} Schultz, P.~H.\ 1996, JGR, 101, 21117

\bibitem[Schultz \& Wrobel(2012)]{schwro2012} Schultz, P.~H., \& Wrobel, K.~E.\ 2012, Journal of Geophysical Research (Planets), 117, E04001

\bibitem[Shoemaker(1962)]{shoemaker1962} Shoemaker, E.~M., 1962, in Physics and Astronomy of the Moon (edited by Z. Kopal), Academic Press, New York and London, pp. 283-359

\bibitem[Soker(1998)]{soker1998} Soker, N.\ 1998, AJ, 116, 1308 

\bibitem[Spiegel \& Madhusudhan(2012)]{spimad2012} Spiegel, D.~S., \& Madhusudhan, N.\ 2012, ApJ, 756, 132 

\bibitem[Stephan et al.(2017)]{steetal2017} Stephan, A.~P., Naoz, S., \& Zuckerman, B.\ 2017, ApJL, 844, L16

\bibitem[Stephan et al.(2018)]{steetal2018} Stephan, A.~P., Naoz, S., \& Gaudi, B.~S.\ 2018, AJ, 156, 128

\bibitem[Stone et al.(2015)]{stoetal2015} Stone, N., Metzger, B.~D., \& Loeb, A.\ 2015, MNRAS, 448, 188 


\bibitem[Sugita \& Schultz(1999)]{sugsch1999} Sugita, S., \& Schultz, P.~H.\ 1999, Journal of Geophysical Research, 104, 30825.

\bibitem[Swan et al.(2019)]{swaetal2019} Swan, A., Farihi, J., Koester, D., et al.\ 2019, MNRAS, 490, 202

\bibitem[Takizawa \& Katsuragi(2019)]{takkat2019} Takizawa, S., \& Katsuragi, H.\ 2019, arXiv:1904.11636

\bibitem[Tsujido et al.(2015)]{tsuetal2015} Tsujido, S., Arakawa, M., Suzuki, A.~I., et al.\ 2015, Icarus, 262, 79.
  
\bibitem[van Lieshout et al.(2014)]{vanetal2014} van Lieshout, R., Min, M., \& Dominik, C.\ 2014, A\&A, 572, A76

\bibitem[van Lieshout et al.(2018)]{vanetal2018} van Lieshout, R., Kral, Q., Charnoz, S., et al.\ 2018, MNRAS, 480, 2784

\bibitem[Vanderbosch et al.(2019)]{vanetal2019} Vanderbosch, Z., Hermes, J.~J., Dennihy, E., et al.\ 2019, submitted to ApJL, arXiv:1908.09839

\bibitem[Vanderburg et al.(2015)]{vanetal2015} Vanderburg, A., Johnson, J.~A., Rappaport, S., et al.\ 2015, Nature, 526, 546 

\bibitem[van Sluijs \& van Eylen(2018)]{vanvan2018} van Sluijs, L., \& Van Eylen, V.\ 2018, MNRAS, 474, 4603

\bibitem[Veras \& Tout(2012)]{vertou2012} Veras, D., \& Tout, C.~A.\ 2012, MNRAS, 422, 1648 

\bibitem[Veras et al.(2013)]{veretal2013} Veras, D., Mustill, A.~J., Bonsor, A., \& Wyatt, M.~C.\ 2013, MNRAS, 431, 1686 

\bibitem[Veras et al.(2014a)]{veretal2014a} Veras, D., Shannon, A., G\"{a}nsicke, B.~T.\ 2014a, MNRAS, 445, 4175 

\bibitem[Veras et al.(2014b)]{veretal2014b} Veras, D., Jacobson, S.~A., G\"{a}nsicke, B.~T.\ 2014b, MNRAS, 445, 2794 

\bibitem[Veras et al.(2014c)]{veretal2014c} Veras, D., Leinhardt, Z.~M., Bonsor, A., G\"{a}nsicke, B.~T.\ 2014c, MNRAS, 445, 2244

\bibitem[Veras \& G\"{a}nsicke(2015)]{vergae2015} Veras, D., G\"{a}nsicke, B.~T.\ 2015, MNRAS, 447, 1049 

\bibitem[Veras et al.(2015a)]{veretal2015a} Veras, D., Eggl, S., G{\"a}nsicke, B.~T.\ 2015a, MNRAS, 451, 2814 

\bibitem[Veras et al.(2015b)]{veretal2015b} Veras, D., Leinhardt, Z.~M., Eggl, S., G{\"a}nsicke, B.~T.\ 2015b, MNRAS, 451, 3453 

\bibitem[Veras et al.(2015c)]{veretal2015c} Veras, D., Eggl, S., \& G{\"a}nsicke, B.~T.\ 2015c, MNRAS, 452, 1945

\bibitem[Veras(2016a)]{veras2016a} Veras, D.\ 2016a, MNRAS, 463, 2958 

\bibitem[Veras(2016b)]{veras2016b} Veras, D.\ 2016b, Royal Society Open Science, 3, 150571

\bibitem[Veras et al.(2016)]{veretal2016} Veras, D., Mustill, A.~J., G{\"a}nsicke, B.~T., et al.\ 2016, MNRAS, 458, 3942 

\bibitem[Veras et al.(2017a)]{veretal2017a} Veras, D., Georgakarakos, N., Dobbs-Dixon, I., \& G{\"a}nsicke, B.~T.\ 2017a, MNRAS, 465, 2053 

\bibitem[Veras et al.(2017b)]{veretal2017b} Veras, D., Carter, P.~J., Leinhardt, Z.~M., \& G{\"a}nsicke, B.~T.\ 2017b, MNRAS, 465, 1008 

\bibitem[Veras et al.(2018a)]{veretal2018a} Veras, D., Georgakarakos, N., G{\"a}nsicke, B.~T., et al.\ 2018a, MNRAS, 481, 2180

\bibitem[Veras et al.(2018b)]{veretal2018b} Veras, D., Armstrong, D.~J., Blake, J.~A., et al.\ 2018b, Astrobiology, 18, 1106

\bibitem[Veras et al.(2019b)]{veretal2019b} Veras, D., Efroimsky, M., Makarov, V.~V., et al.\ 2019b, MNRAS, 486, 3831

\bibitem[Veras et al.(2019a)]{veretal2019a} Veras, D., Higuchi, A., \& Ida, S.\ 2019a, MNRAS, 485, 708

\bibitem[Veras \& Fuller(2019)]{verful2019} Veras, D., \& Fuller, J.\ 2019, MNRAS, 489, 2941

\bibitem[Veras \& Wolszczan(2019)]{verwol2019} Veras, D., \& Wolszczan, A.\ 2019, MNRAS, 488, 153

\bibitem[Veras \& Fuller(2020)]{verful2020} Veras, D., \& Fuller, J.\ 2020, MNRAS 492, 6059   

\bibitem[Veras \& Scheeres(2020)]{versch2020} Veras, D., \& Scheeres, D.~J.\ 2020, MNRAS, 492, 2437

\bibitem[Veras et al.(2020a)]{veretal2020a} Veras, D., McDonald, C.~H., \& Makarov, V.~V.\ 2020a, MNRAS, 269

\bibitem[Veras et al.(2020b)]{veretal2020b} Veras, D., Tremblay, P.-E., Hermes, J.~J., et al.\ 2020b, MNRAS, 268

\bibitem[Vickery(1993)]{vickery1993} Vickery, A.~M.\ 1993, Icarus, 105, 441.

\bibitem[Vickery \& Melosh(1987)]{vicmel1987} Vickery, A.~M., \& Melosh, H.~J.\ 1987, Science, 237, 738.

\bibitem[Villaver \& Livio(2007)]{villiv2007} Villaver, E., \& Livio, M.\ 2007, ApJ, 661, 1192 

\bibitem[Voyatzis et al.(2013)]{voyetal2013} Voyatzis, G., Hadjidemetriou, J.~D., Veras, D., \& Varvoglis, H.\ 2013, MNRAS, 430, 3383 

\bibitem[Wang et al.(2019)]{wanetal2019} Wang, T.-G., Jiang, N., Ge, J., et al.\ 2019, submitted to ApJL, arXiv:1910.04314

\bibitem[Wickramasinghe et al.(2010)]{wicetal2010} Wickramasinghe, D.~T., Farihi, J., Tout, C.~A., Ferrario, L., \& Stancliffe, R.~J.\ 2010, MNRAS, 404, 1984 

\bibitem[Wilson et al.(2014)]{wiletal2014} Wilson, D.~J., G{\"a}nsicke, B.~T., Koester, D., et al.\ 2014, MNRAS, 445, 1878 

\bibitem[Wilson et al.(2015)]{wiletal2015} Wilson, D.~J., G{\"a}nsicke, B.~T., Koester, D., et al.\ 2015, MNRAS, 451, 3237 

\bibitem[Wilson et al.(2016)]{wiletal2016} Wilson, D.~J., G{\"a}nsicke, B.~T., Farihi, J., \& Koester, D.\ 2016, MNRAS, 459, 3282 

\bibitem[Wyatt(2008)]{wyatt2008} Wyatt, M.~C.\ 2008, ARA\&A, 46, 339

\bibitem[Wyatt et al.(2014)]{wyaetal2014} Wyatt, M.~C., Farihi, J., Pringle, J.~E., \& Bonsor, A.\ 2014, MNRAS, 439, 3371 

\bibitem[Xu et al.(2013)]{xuetal2013} Xu, S., Jura, M., Klein, B., Koester, D., \& Zuckerman, B.\ 2013, ApJ, 766, 132 

\bibitem[Xu et al.(2014)]{xuetal2014} Xu, S., Jura, M., Koester, D., Klein, B., \& Zuckerman, B.\ 2014, ApJ, 783, 79 

\bibitem[Xu et al.(2016)]{xuetal2016} Xu, S., Jura, M., Dufour, P., et al.\ 2016, ApJL, 816, L22

\bibitem[Xu et al.(2017)]{xuetal2017} Xu, S., Zuckerman, B., Dufour, P., et al.\ 2017, ApJL, 836, L7 

\bibitem[Xu et al.(2019)]{xuetal2019} Xu, S., Dufour, P., Klein, B., et al.\ 2019, AJ, 158, 242

\bibitem[Yamamoto et al.(2017)]{yametal2017} Yamamoto, S., Hasegawa, S., Suzuki, A.~I., et al.\ 2017, Journal of Geophysical Research (Planets), 122, 1077

\bibitem[Zuckerman et al.(2003)]{zucetal2003} Zuckerman, B., Koester, D., Reid, I.~N., H\"{u}nsch, M.\ 2003, ApJ, 596, 477 

\bibitem[Zuckerman et al.(2010)]{zucetal2010} Zuckerman, B., Melis, C., Klein, B., Koester, D., \& Jura, M.\ 2010, ApJ, 722, 725 

\bibitem[Zuckerman \& Young(2018)]{zucyou2018} Zuckerman, B., \& Young, E.~D.\ 2018, Handbook of Exoplanets, 14

\end{thebibliography}
\end{document}